\documentclass[12pt,preprint]{aastex}
\usepackage{emulateapj5}

\begin{document}

\title{The origin of dust extinction curves 
with or without the 2175 \AA  $\;$ bump in galaxies:
The case of the Magellanic Clouds}

\author{Kenji Bekki} 
\affil{
ICRAR,
M468,
The University of Western Australia
35 Stirling Highway, Crawley
Western Australia, 6009, Australia
}

\author{Hiroyuki Hirashita}
\affil{Institute of Astronomy, and Astrophysics, Academia Sinica, PO Box 23-141,
Taipei 10617, Taiwan}

\and

\author{Takuji Tsujimoto}
\affil{
National Astronomical Observatory, Mitaka-shi, Tokyo 181-8588, Japan}

\begin{abstract}
The Large and Small Magellanic Clouds (LMC and SMC, respectively)
are observed to have characteristic dust extinction curves that are
quite different from those of the Galaxy (e.g., strength of the 2175 \AA  $\;$ bump).
Although the dust composition and size distribution of the Magellanic
Clouds (MCs)
that can self-consistently explain their observed extinction curves
have been already proposed,
it remain unclear whether and how the required dust properties
can be achieved in the formation histories of the MCs.
We therefore 
investigate the time evolution of the dust properties of the MCs
and thereby derive their extinction curves  using
one-zone chemical evolution models with formation and evolution of 
small and large
silicate and carbonaceous dust grains
and dusty winds associated with starburst events.
We find that the observed SMC extinction curve
without a conspicuous 2175 \AA  $\;$ bump  can be reproduced
well by our SMC model, 
if the small carbon grains can be selectively lost through
the dust wind during
the latest starburst about 0.2 Gyr ago. We also find that
the LMC extinction curve with a weak 2175 \AA $\;$ bump
can be reproduced by our LMC model with less efficient removal
of dust through dust wind.
We discuss  possible physical reasons for different dust wind
efficiencies between silicate and graphite and among  galaxies.
\end{abstract}


\section{Introduction}

The Large and Small Magellanic Clouds
(LMC and the SMC respectively), which are the closest pair of interacting
dwarf galaxies around the Milky Way (MW), have long served as 
an ideal laboratory to study many aspects of galaxy formation
and evolution, such as significant influences of galaxy
interaction on galactic star formation 
and chemical enrichment histories
(e.g., Westerlund 1997).
Structures and kinematics (e.g., 
Cole et al. 2005; Nidever et al. 2012),
age and metallicity distributions of stellar populations
(e.g, Da Costa \& Hatzidimitriou 1998; Piatti et al. 2001;
Carrera et al. 2008),
proper motions (e.g., Kallivayalil et al. 2006;  
Cioni et al. 2014),
and gas and dust contents (e.g., Kim et al. 1999;
Meixner et al. 2011; Galliano et al. 2011)
have been extensively investigated so far.
These observed physical
properties of the Magellanic Clouds (MCs) have been used to constrain theoretical models
for the MC evolution, such as the 3D orbits around the MW
and the LMC-SMC interaction histories and their influences on the star formation
histories (e.g., Gardiner \& Noguchi 1995;    
Diaz \& Bekki 2011).

One of the long-standing problems related to the physical
properties of the MCs is the origin
of their extinction curves that are quite different from those
of the MW (e.g., Fitzpatrick 1989). 
The MW extinction curve clearly shows the 2175 \AA $\,$ feature (``bump")
whereas the 2175 \AA $\,$ bump  is almost absent in the SMC and rather weak
in the LMC (e.g., Savage \& Mathis 1979;
Rocca-Volmerange et al. 1981; Prevot et al. 1984; Clayton \& Martin 1985;
Fitzpatrick 1985; Gordon \&  Clayton 1998; Gordon et al. 2003). 
Furthermore, the extinction curves of the MCs at shorter wavelengths
($1/\lambda > 5 \mu {\rm m}^{-1}$) are different from those of the MW
in the sense that they more steeply rise (e.g., Hutchings 1982;
Nandy et al. 1982; Bromage \& Nandy 1983).
These observations suggest that the optical and physical properties
of dust grains (e.g., size distributions) in the MCs are quite
different from those of the MW for some physical reason.

Pei (1992) adopted the graphite-silicate grain model for the MCs
and demonstrated that the observed extinction curves can be reproduced
reasonably well by  the models with different abundance ratios of graphite to
silicate. Weigartmer \& Draine (2001) successfully reproduced
the observed extinction curves of the MCs and the MW by adopting
a simple functional form for the size distributions of carbonaceous
and silicate grains. Although these previous studies clarified
dust compositions and size distributions
required for explaining the observed extinction curves of the MCs,
it remains unclear whether and how the required dust properties
can be achieved in the formation histories of the MCs.
Since the time  evolution of dust properties depends on star formation
and chemical enrichment histories of galaxies (e.g., Dwek 1998),
a dust evolution model that can explain self-consistently
the observed SFH and chemical abundances of the MCs needs to be constructed
for better understanding the origin of the extinction curves of the MCs.

Asano et al. (2013) have first incorporated the time evolution
of dust size distributions into one-zone chemical evolution models
in a fully self-consistent manner
and thereby discussed  what determines the dust size evolution
in galaxies. 
Nozawa et al. (2015) have additionally incorporated
the molecular cloud phase hosting quick dust growth 
and thereby shown that the observed peculiar extinction curves of high-redshift
quasars ($z\ge 4$) can be reproduced,  if the quasars can have
both a rather high molecular gas fraction ($\ge 0.5$) 
and amorphous carbon instead of graphite.
These studies, however, have not attempted to reproduce  the observed
characteristic extinction curves with the nearly absent 2175 \AA $\;$
feature in the SMC and the weak one in the LMC.
Although recent numerical simulations of  the MCs have shown the time evolution
and the spatial distributions
of dust abundances in the MCs (Yozin \& Bekki 2014), they did not
show the size distribution  of dust grains in the MCs. Thus, it remains 
theoretically unclear in what physical conditions 
the observed extinction curves of the MCs can be reproduced
in the evolution models of the MCs.

The purpose of this paper is to investigate the
dust extinction curves of the MCs by using new one-zone chemical evolution
models with the evolution of dust abundances, compositions, and size
distributions.
In particular, we investigate in what physical conditions
the observed characteristic extinction 
curves can be reproduced in the new models.
Since star formation histories (SFHs), age-metallicity relations
(AMRs), and chemical properties (e.g., [Fe/H] and [$\alpha$/Fe])
of the MCs have been derived (e.g., Harris \&   Zaritsky 2009;
Rubele et al. 2012),
we need to construct chemical evolution models that are consistent
with these observations. Furthermore, recent observational studies
by infrared space telescopes, Akari, Spitzer,  and Herschel,
have revealed a number of intriguing dust properties in the MCs
(e.g., Meixner et al. 2010).
These observations can be used to constrain the present
dust evolution models for the MCs (e.g., Galliano et al. 2011).

The plan of the paper is as follows: In the next section,
we describe our new one-zone chemical evolution models
which include not only some key dust-related physical processes
(e.g., growth and destruction) but also radiation-driven dust wind
depending on dust grains.
In \S 3, we
present the results
on the time evolution of the extinction curves for the representative
SMC, LMC, and MW models.
In \S 4, we discuss (i) the validity of the present new scenario that
explains the origin of the observed extinction curves of the MCs
and (ii) alternative ones.
We do not discuss the details of
the chemical evolution of the SMC and the LMC 
and their dependences on the evolution histories
of the MCs in the present
paper, because they have been already discussed in our previous papers
(e.g., Tsujimoto \& Bekki 2009;  Bekki \& Tsujimoto 2010, 2012,
BT12).

Although the formation process of dust in supernovae (SNe) and 
Asymptotic Giant Branch (AGB) stars
in the MCs
have been investigated both observationally and theoretically
(e.g., Matsuura et al. 2011; Zhukovska \& Henning 2013;
Schneider et al. 2014), we do not discuss these issues in the present
paper. The observed spatial distributions of dust properties 
in the MCs (e.g., Galliano et al. 2011) are not discussed either
in the present study, because they are beyond the
scope of this paper. The present one-zone models do not allow
us to investigate the observed spatial variations in the
extinction curves of the MCs (e.g., Gordon \& Clayton 1998),
and we will discuss this important issue in our future papers in which
hydrodynamical simulations of the MCs with a model for dust size
distributions will allow for such an investigation of the spatial
variation of the extinction curves.

\section{The model}

\subsection{Outline}

We adopt the following two steps to derive the dust extinction curves
of the MCs. First, we investigate the time evolution of
dust grains by using one-zone chemical evolution models of the MCs
that have been developed by BT12.
Second, we derive the extinction curves based on the abundances
of dust grains by using the ``two-size approximation model''
developed by Hirashita (2015; H15).
In this two-size approximation, the whole grain population is represented
by small grains (radius $a$ less than 0.03 $\mu$m) and large grains
($a > 0.03$ $\mu$m), because the major grain processing mechanisms
work differently on small and large grains (H15).
The original purpose of this two-size approximation method is to use it
in statistical (theoretical) predictions
and Nbody+hydrodynamical simulations of galaxy formation and evolution
and thereby to  dramatically reduce the total amount of
calculation time for estimating the dust extinction curve for each galaxy
and each particle of a simulated galaxy.
We use the method in the present study, because we need to run numerous models
and the method enables us to complete such calculations of extinction curves
much faster.

In the first step,
each element (e.g., C, O, and Si) is divided into
gas-phase metals and dust,
and dust is further divided into ``small'' and ``large'' grains.
We accordingly investigate the time evolution of the three components,
gas-phase metals
and small and large grains,  separately for each element (e.g., C, O, and Si)
in a model. The time evolution of small and large grains can be used
for the derivation of the extinction curve  in 
the second step.
As demonstrated by H15,
the two-size approximation for a dust size distribution is very
useful and convenient, because a full calculation of the dust size
distribution in a galaxy is complicated and time-consuming
(e.g., Asano et al. 2013). 
Accordingly we adopt the revised version of the two-size approximation model
(H15) in order to derive the extinction curve of a galaxy
in the second step.
In the revised version of the two-size approximation model,
the dust in the ISM of a galaxy is divided into four grains;
small and large silicate (referred to as
small and large  ``Si-grains'', respectively)
and small and large carbonaceous grains (small and large  ``C-grains'',
respectively). This two-size, two-component dust grain model enables
us to derive the extinction curve of a galaxy based on the
chemical evolution of the galaxy, as described later.

Star formation histories (SFHs) of galaxies can control their chemical evolution
histories, because chemical evolution depends on the time evolution
of metals ejected from SNe and AGB stars, the formation
rates of which depend on star formation rates (SFRs).
Therefore,  the  dust evolution histories of galaxies, which depend
on their chemical evolution histories,  can be controlled by 
their SFHs. 
Since we mainly discuss the extinction curves of the MCs,
we need to choose carefully the model parameters for star formation
of the MCs. Many observations suggest that the MCs have experienced
at least a few starburst (SB) events in the past 6 Gyrs 
(e.g., Harris \& Zaritsky 2004, 2009;  Rubele et al.  2012, 2015). 
Furthermore,  previous and recent numerical simulations of the MCs
have demonstrated that highly enhanced star formation is inevitable
during the past LMC-SMC-Galaxy tidal interaction
(e.g., Yoshizawa \& Noguchi 2003; Bekki \& Chiba 2005; Yozin \& Bekki 2014).
We consider these observational
and theoretical results in constructing the SFHs of the MCs.

Since the importance and possibility 
of dust removal from galaxies through energetic radiation-driven
dust wind in one-zone chemical evolution models have
 been discussed extensively by Bekki \& Tsujimoto (2014, BT14),
we do not discuss this issue in detail in this paper.
The present model is more sophisticated than BT14 in the sense that
it investigates the evolution of gas-phase metals
and the formation and evolution of different dust grains
in a self-consistent manner.
Furthermore, the present model newly includes,
the accretion of gas-phase metals onto already existing dust grains,
its dependence on cold gas fraction in the ISM,
grain growth by coagulation,
and grain disruption by shattering.
Thanks to this, we can provide more reliable predictions on  the time
evolution of dust grains in a galaxy.

\subsection{Basic equations}

We adopt a revised version of our one-zone chemical evolution
models used in BT12 in order to investigate the time evolution
of gas, metals, and dust in a fully self-consistent manner.
Although there are only two basic equations in BT12,
there are four in the present study,
which reflects the fact that the revised version
includes the evolution of small and large dust grains.
We investigate the time evolution of the total masses  of gas ($M_{\rm g}$)
and that of the total masses of 
metal ($M_{\rm z}$) and dust ($M_{\rm d}$) for each element (e.g., C and O)
in a model. Since the total metal mass includes gas-phase
metals and dust,  the total mass of {\it gas-phase metals}
that is not depleted onto dust grains  ($M_{\rm z,g}$)
can be  estimated as $M_{\rm z}-M_{\rm d}$ in the present study.
The total gas mass ($M_{\rm g, t}$) accreted onto a galaxy
is set to be 1.0 (i.e., normalized) in all models of the present study.
The basic equations for the time evolution
of gas and metals in the adopted one-zone chemical evolution models
with dust physics are described as follows:
\begin{eqnarray}
\frac{dM_{\rm g}}{dt}=-\alpha_{\rm l}\psi(t)+A(t)-w(t)-w_{\rm d}(t)
\end{eqnarray}

\begin{eqnarray}
\frac{dM_{\rm z, \it i} }{dt}=-\alpha_{\rm l} Z_i(t)\psi(t)
+Z_{A,i}(t)A(t)
+y_{{\rm II},i}\psi(t)
\nonumber \\  
 +y_{{\rm Ia},i}\int^t_0
\psi(t-t_{\rm Ia})g(t_{\rm Ia})dt_{\rm Ia}
\nonumber \\
 +\int^t_0 
y_{{\rm agb},i}(m_{\rm agb})
\psi(t-t_{\rm agb})h(t_{\rm agb})dt_{\rm agb} 
\nonumber \\
-W_i(t) -W_{\rm d, \it i}(t) \ \ ,
\end{eqnarray}
\noindent where $\alpha_{\rm l}$ is the mass fraction
locked up in dead stellar remnants and long-lived stars,
$\psi(t)$ is the SFR,
$Z_i$ is the metallicity for $i$th element,
$y_{\rm {II}, i}$, 
$y_{\rm {Ia}, i}$, 
and $y_{\rm {agb}, i}$ are the
chemical yields for the $i$th element from type II supernovae (SNe II),
from Type Ia supernovae (SNe Ia),
and from AGB stars, respectively,
$Z_{A,i}$ is the abundance of heavy elements  contained in the infalling gas,
and $W_i$ is the wind rate for each element.
The quantities $t_{\rm Ia}$ and $t_{\rm agb}$ represent
the time delay between star formation and SN Ia explosion
and that between star formation and the onset of AGB phase,
respectively.
The terms $g(t_{\rm Ia}$) and $h_{\rm agb}$ are the distribution
functions of SNe Ia and AGB stars, respectively,
and the details of which  are described later
in this section.
The term $h_{\rm agb}$ controls
how much AGB ejecta can be returned into the ISM per unit mass for a given time
in equation (2).
The total gas masses  ejected from AGB stars depend on the original masses
of the AGB stars (e.g., Weidemann 2000).
Therefore, this term  $h_{\rm agb}$ depends on
the adopted IMF and the relation between the mass and the lifetime
in stars  (later described).
The term  $W_{\rm d, \it i}$ is the dust wind rate for each element
(i.e., $w_{\rm d}$ is the sum of $W_{\rm d, \it i}$ for all elements).

The basic equations for the time evolution
of the total masses of 
small ($M_{\rm d, s}$) and large ($M_{\rm d, l}$) grains
in the  models
are described as follows:
\begin{eqnarray}
\frac{dM_{\rm d,s,\it i}}{dt}=
-\alpha_{\rm l}D_{\rm s,\it i}\psi(t)
+D_{\rm s,\it i}(0)A(t)
-\frac{ M_{\rm d,s,\it i} }{ \tau_{\rm SN,s, \it i} } 
\nonumber \\
+\frac{ M_{\rm d,l,\it i} }{ \tau_{\rm sh,\it i} } 
-\frac{ M_{\rm d,s,\it i} }{ \tau_{\rm co,\it i} } 
+\frac{ M_{\rm d,s,\it i} }{ \tau_{\rm acc,s,\it i} } 
-W_{\rm d,s,i}(t)
\end{eqnarray}

\begin{eqnarray}
\frac{dM_{\rm d,l,\it i}}{dt}=
-\alpha_{\rm l}D_{\rm l,\it i}\psi(t)
+D_{\rm l,\it i}(0)A(t)
+f_{\rm c, \it i} E_{\rm z, \it i}
\nonumber \\
-\frac{ M_{\rm d,l,\it i} }{ \tau_{\rm SN,l, \it i} } 
-\frac{ M_{\rm d,l,\it i} }{ \tau_{\rm sh,\it i} } 
+\frac{ M_{\rm d,s,\it i} }{ \tau_{\rm co,\it i} } 
+\frac{ M_{\rm d,l,\it i} }{ \tau_{\rm acc,l,\it i} } 
-W_{\rm d,l,\it i}(t)
\end{eqnarray}

\noindent where 
$M_{\rm d,s, \it i}$ and
$M_{\rm d,l, \it i}$ 
are  the masses of small and large  grains for
$i$-th element, respectively,
$D_{\rm s, \it i}$ and
$D_{\rm l, \it i}$  
are the dust-to-gas-ratios for small and large grains, respectively,
$D_{\rm s, \it i}$(0) and
$D_{\rm l, \it i}$(0)  
are the initial dust-to-gas-ratios for small and large grains
(in infalling gas), respectively,
$f_{\rm c, \it i}$ and $E_{\rm z, \it i}$
are  the condensation efficiency of dust and the ejection rate of a metal
element, respectively,
$\tau_{\rm SN, s, \it i}$ and 
$\tau_{\rm SN, l, \it i}$ 
are the timescales of dust destruction by
SNe for small and large grains, respectively,
$\tau_{\rm sh, \it i}$ is the timescale for shattering of large grains,
$\tau_{\rm co, \it i}$ is the timescale for coagulation of small grains,
and $W_{\rm d, s, \it i}$ and 
$W_{\rm d, l, \it i}$ 
are the dust wind rates
for small and large grains, respectively.
The dust wind term $W_{\rm d, \it i}$ in equation (2)
is therefore equivalent to $W_{\rm d, s \it i} + W_{\rm d, l, \it i}$.

The  ejection rate of metals from SNe and AGB stars
($E_{\rm z, \it i}$) is described as follows:
\begin{eqnarray}
E_{\rm z, \it i}=y_{{\rm II},
i}\psi(t)
+y_{{\rm Ia},i}\int^t_0
\psi(t-t_{\rm Ia})g(t_{\rm Ia})dt_{\rm Ia}
\nonumber \\
+\int^t_0
y_{{\rm agb},i}(m_{\rm agb})
\psi(t-t_{\rm agb})h(t_{\rm agb})dt_{\rm agb}.
\ \
\end{eqnarray}
The formation rate of dust grains from SNe and AGB ejecta at 
each time step is estimated
from $E_{\rm z, \it i}$ and $f_{\rm c, \it i}$. 
It is assumed here that the dust produced by stars
is in the large-grain regime (H15).
It should be also stressed that the dust condensation efficiency for each
$i$th element is different between SN Ia, SNII, and AGB ejecta and
this difference is properly included in the present one-zone models:
the way to describe the
dust production term in  equation (4) is not very accurate,
and such description is adopted only for convenience.
In the implementation of the dust production
process,  the dust production term is calculated in our code as follows:
\begin{eqnarray}
f_{\rm c, \it i}E_{\rm z, \it i} = f_{\rm c, SNIa, \it i}E_{\rm z, SNIa, \it i}
+f_{\rm c, SNII, \it i}E_{\rm z, SNII, \it i}
\nonumber \\
+f_{\rm c, AGB, \it i}E_{\rm z, AGB, \it i},
\end{eqnarray}
where 
$f_{\rm c, SNIa,  \it i}$
$f_{\rm c, SNII,  \it i}$
and $f_{\rm c, AGB,  \it i}$
are the dust condensation efficiencies for $i$th element
and 
$E_{\rm z, SNIa,  \it i}$
$E_{\rm z, SNII,  \it i}$
and $E_{\rm z, AGB,  \it i}$
are the metal ejection rates for $i$th element.

It should be noted here that the way to describe the evolution
of gas, metals, and dust in the above  equations is slightly
different from those adopted in BT12 and H15. For example, 
we investigate  $M_{\rm g}$ and $M_{\rm z}$
instead of the gas mass fraction ($f_{\rm g}$ in BT12)
and metallicity ($Z_i$)  
in the present study.
Also, the timescales for dust accretion, coagulation, and shattering
are defined for each element in the present study: they are assumed
to be constant for all elements in H15. 
The dust wind for small and large grains is separately
considered, for the first time,
in the present study.
The details of the models for star formation, chemical evolution,
and dust physics in the ISM and the model
parameters are described in the following subsections.

\subsection{Star formation and gas accretion}

The star formation rate $\psi(t)$ is assumed to be proportional
to the gas mass with a constant star formation
coefficient and thus is described as follows:
\begin{equation}
\psi(t)=C_{\rm sf}M_{\rm g}(t) .
\end{equation}
The normalization factor $C_{\rm sf}$ is chosen appropriately so that
the final gas mass fraction (and metallicity) can be consistent with
the observed one.
We assume that $C_{\rm sf}$ is different between (i) the ``quiescent phase'',
when a galaxy  shows an
almost steady star formation, and (ii) the ``star burst (SB) phase'',
when the galaxy experiences a bursty  star formation. 
Many observational studies showed that the LMC and the SMC have experienced
a few SB events in the past 6 Gyr (e.g.,  Harris \& Zaritsky 2006, 2009;
Rubele et al. 2012).
Guided by these observations,
we investigate models 
in which a starburst can occur 
three times  (``SB models''), and the first, second, third starbursts
are denoted as SB1, SB2, and SB3, respectively.
The star formation rate  is assumed to be constantly higher for
$t_{\rm sb1,s} \le t \le t_{\rm sb1,e}$ in SB1,
$t_{\rm sb2,s} \le t \le t_{\rm sb2,e}$ in SB2,
and $t_{\rm sb3,s} \le t \le t_{\rm sb3,e}$ in SB3.
The star formation coefficient  ($C_{\rm sf}$) is thus described as
follows:

\begin{equation}
C_{\rm sf}= \left\{
\begin{array}{lll}
C_q & \mbox{for quiescent phase} \\
C_{\rm sb1} & \mbox{for SB1} \\
C_{\rm sb2}  & \mbox{for SB2} \\
C_{\rm sb3}  & \mbox{for SB3} \\
\end{array}
\right.
\end{equation}
In the present study,  $C_{\rm sf}$ is given
for each model (e.g., SMC, LMC, and MW).

We adopt the following formula for the rate of gas accretion
onto the disk of a galaxy:
\begin{equation}
A(t)=C_{\rm a}\exp(-t/t_{\rm a}),
\end{equation}
where $t_{\rm a}$ is a free parameter controlling the timescale of
the gas accretion and $C_{\rm a}$ is the normalization
factor and determined such
that the total gas mass ($M_{\rm g, t}$)
 accreted onto the galaxy can be 1 for a given
$t_{\rm a}$. This $M_{\rm g, t}$ is determined such that
$M_{\rm g, t}$ is 
the total baryonic mass of a (present) galaxy (i.e.,
gas and stars, {\it including gas , metals,
and dust ejected from the  galaxy through outflow}).
The parameter $t_{\rm a}$ can therefore control the time evolution
of the total gas mass and the gas mass fraction of a galaxy.
We mainly show the results of the models
with $t_{\rm a}=5$ Gyr,
though we investigated  models with different $t_{\rm a}$. 
The main purpose is to discuss how dust wind plays  key roles
in the evolution of extinction curves of galaxies in the present study.
Such roles of dust wind do not depend on the adopted $t_{\rm a}$.

Although we show only the results of the models with $t_{\rm a}=5$ Gyr,
the final (i.e., present) gas masses and gas mass fractions 
of the three galaxies  are roughly consistent with the observed values.
For example, if we adopt $M_{\rm g}=3 \times 10^9 M_{\odot}$ (BT12) for the LMC,
then the final gas mass is $1.1 \times 10^9 M_{\odot}$ 
and the gas fraction is 0.37. These numbers are roughly consistent with 
the observed values (e.g., van den Bergh 2000), which suggests that
the adopted $t_{\rm a}$ is reasonable.
Furthermore 
the  typical star formation rate at non-starburst phases in the LMC models is 
0.15 $M_{\odot}$ yr$^{-1}$, which is similar to the observed present SFR of the LMC
($0.26 M_{\odot}$ yr$^{-1}$; Kennicutt et al. 1995).
The final gas mass fractions of the SMC and the MW are 0.53 and 0.05 in the present 
models,  respectively,
which are also roughly consistent with the observed values (e.g., van den Bergh 2000).
The evolution of SFRs and gas mass fractions are given in Figure 1 and later
discussed.

The assumption of slow gas accretion onto the SMC implies that the SMC has a disk
component.
Almost all previous theoretical models of the Magellanic Stream (MS) formation needed to
assume the gas and stellar disks (e.g., Gardiner \& Noguchi 1995; Diaz \& Bekki
2012) in order to reproduce quite well the observed properties of the MS.
Furthermore, recent observations have confirmed that the SMC has a stellar disk
(e.g., Dobbie et al. 2014), and it is a well known fact that the SMC has a rotating
HI disk (e.g., Stanmirovic et al. 2004).
Therefore, it is reasonable for the present study to assume that
the SMC has a disk onto which gas can be accreted.

\subsection{SN-driven wind}

We assume that  some fractions of metals  from  SNe
can be removed from a galaxy thorough
energetic outflows so that they do not contribute to
chemical enrichment  of the galaxy.
In the present study
only SNe ejecta can be expelled from
a galaxy (``selective wind model'').
We do not consider that the ejection efficiency  depends
on (i) chemical elements (e.g., C and O) and (ii) whether the ejecta
is from SN Ia or SN II  just for simplicity in the present study.
Some fraction, $1-f_{\rm ej}$,
of gaseous ejecta from SNe can be mixed with the ISM for chemical enrichment
processes whereas all AGB ejecta can be mixed with the ISM
in the selective wind models.

The SN wind rate ($w(t)$) is estimated only from the total
mass of gaseous ejecta from SNe ($M_{\rm ej, sn}$)
at each time step.  The total mass, $M_{\rm w}$, of ISM that is removed from
a galaxy  at each time step is as follows:
\begin{equation}
M_{\rm w}=f_{\rm ej}M_{\rm ej, sn} .
\end{equation}
Therefore, the  SN wind rate at each time step is simply
\begin{equation}
w(t)=\frac{dM_{\rm w} }{dt}.
\end{equation}
We show the results of the models with $f_{\rm ej}=0.4$ in the present study,
because BT12 have already investigated the SN wind models with $f_{\rm ej}=0.2$ 
and 0.4 and confirmed that such models can better reproduce the observed properties
of chemical abundances of the LMC (e.g., Figure 16 of BT12). 
We also confirm that the roles of dust wind in the evolution of extinction curves
do not depend strongly on $f_{\rm ej}$, though the models with lower $f_{\rm ej}$
end up with higher gaseous metallicities.

Multiple SN events can strongly influence the chemical and dynamical evolution
of star-forming regions, and one of the most dramatical influences is the formation
of super shells and bubbles in the H~{\sc i} disks of the LMC and the SMC
(e.g., Kim et al. 1999; Stanimirovic et al. 1999). Such SN events can cause more
efficient dust destruction, shattering, and coagulation in shells and bubbles
of  MCs, which implies
that the models of dust evolution would need to include such dust-related physical
processes depending on the time evolution of SN rates. In the present study, we
do not intend to model  such dust-related physical processes dependent on
SN rates, mainly because such complicated models require  a few
additional model  parameters and thus would possibly prevent us from
clarifying the key parameters for the evolution of dust extinction curves.
Thus it would be our future study to include the dust-related physical processes
dependent on SN rates of galaxies in dust evolution models. 

\subsection{Delay time distribution of SN Ia}

We assume that there is a time delay ($t_{\rm Ia}$) between the star formation
and the metal ejection from SNe Ia. We here adopt the following 
delay time distribution, ``DTD'',
($g(t_{\rm Ia}$)) for 0.1 Gyr $\le t_{\rm Ia} \le$ 10 Gyr,
which is deduced from recent observational studies
on the SN Ia rate in extra-galaxies (e.g., Maoz et al. 2010);
\begin{equation}
g_{\rm Ia} (t_{\rm Ia})  = C_{\rm g}t_{\rm Ia}^{-1},
\end{equation}
where $C_{\rm g}$ is a normalization constant that is determined by
the number of SN Ia per unit mass  (which is controlled by the IMF
and the binary fraction for intermediate-mass stars
for  the adopted power-law slope of $-1$).
We mainly investigate models 
with
the above DTD (rather than those 
with typical time delay of SN Ia being  $\sim 1$ Gyr).
The fraction of the stars that eventually
produce SNe Ia for 3--8$M_{\odot}$ has not been observationally determined
and thus is regarded as a free parameter, $f_{\rm b}$.
We investigate models with $f_{\rm b}=0.05$,
because such models can better explain the observed chemical properties
of the MCs and the Galaxy (e.g., Tsujimoto et al. 2010; BT12).

\subsection{Nucleosynthesis and dust yields}

We adopt  the nucleosynthesis yields of SNe II and Ia from T95
to deduce $y_{{\rm II},i}$ and $y_{{\rm Ia},i}$ for a given IMF.
For AGB yields,
we calculate $y_{\rm agb, \it i}$ and $h(t_{\rm agb})$ for a given
IMF based on the yield tables 
from van den Hoek \& Groenewegen (1997).
Since our recent chemodynamical simulations of disk galaxy formation
can reproduce the basic dust properties of galaxies (e.g., dust-to-gas-ratios)
reasonably well (Bekki 2013), we adopt the dust yields
(corresponding  to $f_{\rm c, \it i}$)  used in 
the simulations for the present study.
The adopted $f_{\rm c, \it i}$ is originally from  Dwek (1998), which
explained the dust properties of the Galaxy quite well.
The adopted values are different between different elements
and between SNe and AGB stars. 
We investigate the time evolution of
dust and metals for the selected eight elements, 
C, O, Mg, Fe, Ca, SI, Ti, and S.

\subsection{IMF}

We adopt a standard power-law IMF;
\begin{equation}
\Psi (m_{\rm I}) = M_{s,0}{m_{\rm I}}^{-\alpha},
\end{equation}
where $m_{\rm I}$ is the initial mass of
each individual star and the slope $\alpha$ is set to be 2.35, which 
corresponds to the Salpeter IMF  (Salpeter 1955).
The normalization factor $M_{s,0}$ is a function of $\alpha$,
$m_{\rm l}$ (lower mass cut-off), and  $m_{\rm u}$ (upper mass cut-off).
These  $m_{\rm l}$ and $m_{\rm u}$
are  set to be   $0.1 {\rm M}_{\odot}$
and  $50 {\rm M}_{\odot}$, respectively.
As discussed in BT12, 
the IMF slope $\alpha$ can influence the chemical evolution history
of the LMC. We however did not discuss such an influence in the present
study, firstly  because the purpose of this paper is not to clarify the
roles of IMF in galactic chemical evolution and secondly because
the models with the standard IMF (with SN wind) can explain the 
observed abundances of the MCs (e.g., Tsujimoto \& Bekki 2009; BT12).

The IMF slope can possibly influence the evolution of dust and metals
of galaxies, because it can influence the total amount of energy ejected
from massive stars and SNe. If we adopt a top-heavy IMF (e.g., $\alpha=2.1$ instead
of 2.35), then
dust and metals can be more efficiently ejected from the MCs so that dust-to-gas-ratios
can be lower in the MCs. The time evolution of extinction curves of the MCs might
be changed to some extent owing to the more efficient removal of dust. Although
this is an important issue, we do not investigate how the IMF can influence
the evolution of dust and extinction curves of the MCs in the present study:
It is our future study to clarify these IMF influences.

\subsection{Two-composition approximation}

Although we investigate the time evolution of small and large grains
for each element (e.g., C, O, and Mg) {\it separately},
we divide these grains into the following two categories: C-grains and
Si-grains. Dust grains formed from the element C are referred to as C-grains
whereas those from elements other than C are referred to as Si-grains.
Asano et al. (2013) also adopted this treatment of Si-grains, expecting that
solid materials other than carbonaceous dust compose silicate
materials after dust processing in the ISM.
Accordingly, there are four different types of grains
({\it four-component dust model}), for each of which
the total mass and dust-to-gas-ratio ($D$) are 
investigated in the present study.
For example, the dust-to-gas-ratio for small Si-grains is estimated  as follows:
\begin{equation}
D_{\rm s,Si}= \sum_i D_{\rm s, i},
\end{equation}
where the summation is done for elements 
other than C (i.e., only for $i$=O, Fe, Mg, Ca, Si, Ti, and S).
The dust-to-gas-ratio
for small C-grains is simply
$D_{\rm s, C}$. Thus, the total dust-to-gas-ratio ($D$) is given as:
\begin{equation}
D= \sum_i \sum_j D_{i,j},
\end{equation}
where summation is done over $i$=s (small) and l (large) and
$j$=C, O, Fe, Mg, Ca,
Si, Ti, and S.

\subsection{Dust growth}

We consider (i) that dust grains can grow quite efficiently through the
accretion of gas-phase metals onto the grains in the cold 
molecular clouds, (ii) that  the timescale is shorter for clouds with higher
metal-abundance,  and (iii) that  dust accretion timescale
($\tau_{\rm acc, s, \it i}$ and $\tau_{\rm acc, l, \it i}$)
depends on the metallicity and the molecular gas density in ISM.
The accretion timescale for the small grain of
$i$th element is therefore as follows:
\begin{equation}
\tau_{\rm acc, s, \it i} = \frac{ \tau_{\rm acc, s, \it i, \odot} }
{ C_{\rm acc, \it i } },
\end{equation}
where $\tau_{\rm acc, s, \it i, \odot}$ is the dust accretion timescale
at the solar metallicity (0.014) and the total dust abundance (0.0064)
at the solar neighborhood (in the present MW),
and the correction factor $C_{\rm acc, \it i}$ is as follows:

\begin{equation}
C_{\rm acc, \it i } = F(Z_{\rm z, g}, f_{\rm H_2})
=\frac{ Z_{\rm z, g} }{ Z_{\rm z,g, \odot} }
( \frac{ D }{ D_{\odot}} )^{\beta}, 
\end{equation}
where $D$, $Z_{\rm z, g}$, $f_{\rm H_2}$  are the dust-to-gas-ratio,
the gas-phase metallicity, and the cold gas (molecular gas)
fraction, respectively, and $Z_{\rm z,g, \odot}$ 
and $D_{\odot}$ are 
the gas-phase metallicity and 
the dust-to-gas-ratio at the solar neighborhood, respectively.
Here, $D_{\odot}$ is
the total dust-to-gas-ratio and thus set to be 0.0064,
and $\beta$ relates the dust-to-gas-ratio to the molecular fraction
of ISM. 
In this formula,  we adopt a very reasonable assumption 
that the dust accretion timescale
is shorter in higher metallicity and higher cold gas fraction.
The cold gas fraction 
is  assumed to be a function of  $D$ as follows:
\begin{equation}
f_{\rm H_2} \propto
( \frac{ D }{ D_{\odot}} )^{\beta}, 
\end{equation}
where we estimate $f_{\rm H_2}$ by using $D$ and gas mass
at each time step in a model.
We choose this $\beta$ of 0.2 for $f_{\rm H_2}$, 
because such a choice  is consistent
with recent observational results by Corbelli et al. (2012) who show
a correlation between the total masses of molecular hydrogen and dust in
galaxies.

Although we  adopt the exactly the same formula for
the dust accretion timescale 
($\tau_{\rm acc, l, \it i}$)
of large grains,
we consider that 
$\tau_{\rm acc, l, \it i, \odot}$ should be significantly
longer than
$\tau_{\rm acc, s, \it i, \odot}$ owing to the possible
mass growth timescale proportional to dust sizes (H15).
We therefore assume that $\tau_{\rm acc, l, \it i, \odot}$ is 
$10\tau_{\rm acc, s, \it i,\odot}$
for all models. 
We show the results of the models with
$\tau_{\rm acc, s, \it i, \odot}=0.02$ Gyr
$\tau_{\rm acc, l, \it i, \odot}=0.2$ Gyr for all elements
in the present study.
The dust destruction timescale 
($\tau_{\rm SN, s, \it i}$
and $\tau_{\rm SN, l, \it i}$)
is set to be 0.2 Gyr for small and large grains of all elements,
and these values are similar to those adopted in previous studies (e.g.,
Dwek 1998; H15).

\subsection{Coagulation and shattering}

As discussed by H15, the timescales of coagulation and shattering
depend on $D$ (dust-to-gas-ratio) in the ISM of galaxies. We therefore
adopt the following formula:
\begin{equation}
\tau_{\rm sh,  \it i}=\tau_{\rm sh,  \it i, \odot}
( \frac{ D_{\rm l  \it i} }{ D_{\rm l, \it i ,\odot} } )^{-1},
\end{equation}
where 
$\tau_{\rm sh,  \it i, \odot}$ 
and $D_{\rm l, \it i, \odot}$ are
$\tau_{\rm sh,  \it i}$ 
and $D_{\rm l, \it i}$ (dust-to-gas-ratio for
large grains) 
at the solar neighborhood, respectively.
Since C (carbon) becomes carbon dust whereas metals other
than C become silicate,  
the shattering timescale for large C-grains is as follows:
\begin{equation}
\tau_{\rm sh,  \it i}=\tau_{\rm sh,  C, \odot}
( \frac{ D_{\rm l, C} }{ D_{\rm l, \odot} } )^{-1},
\end{equation}
where $D_{\rm l, C}$ is the dust-to-gas-ratio of large C-grains
and $D_{\rm l, \odot}$ is the dust-to-gas-ratio for
large grains (0.0045) at the solar neighborhood.
The timescale  for elements other than C is as follows:
\begin{equation}
\tau_{\rm sh,  \it i}=\tau_{\rm sh,  Si, \odot}
( \frac{ D_{\rm  l, Si } }{ D_{\rm  l, \odot} } )^{-1},
\end{equation}
where $D_ {\rm l, Si}$ is the dust-to-gas-ratio of large C-grains.
We choose $\tau_{\rm sh,  C, \odot}= 
\tau_{\rm sh, Si}=0.1$ Gyr, which is similar to those adopted
in previous studies (e.g., H15).

Similarly, the coagulation timescale for C is described as follows:
\begin{equation}
\tau_{\rm co,  \it i}=\tau_{\rm co,  C, \odot}
( \frac{ D_{\rm s, C} }{ D_{\rm s, \odot} } )^{-1},
\end{equation}
where $D_{\rm l, C}$ is the dust-to-gas-ratio of small C-grains
and $D_{\rm l, \odot}$ is the dust-to-gas-ratio for
small grains (0.0019) at the solar neighborhood.
The coagulation timescale for elements other C  is  as follows:
\begin{equation}
\tau_{\rm co,  \it i}=\tau_{\rm co,  Si, \odot}
( \frac{ D_{\rm s, Si} }{ D_{\rm s, \odot} } )^{-1},
\end{equation}
where $D_{\rm l, Si}$ is the dust-to-gas-ratio of small Si-grains.
The values of 
$\tau_{\rm co,  C, \odot}$ 
and $\tau_{\rm co,  Si, \odot}$ are set to be 0.02 Gyr, which is similar
to those adopted by other studies (e.g., H15).

\subsection{Radiation-driven dust wind}

The new model for dust wind rate $W_{\rm d}(t)$
in the present study is different to some extent from
those adopted in our previous study (BT14).
The wind rate is assumed to be different not only 
between C- and Si-grains but also between small and large grains
in the new model.
Therefore the wind rate $W_{\rm d}(t)$ is described as follows:
\begin{equation}
W_{\rm d}(t)=C_{\rm w}  S(t) M_{\rm d}(t),
\end{equation}
where $C_{\rm w}$ is a parameter that controls the removal efficiency
of dust through dust wind, $S(t)$ is the total luminosity
(proportional to the strength of radiation pressure
of stars), and $M_{\rm d}$ is the total dust mass.
The physical meaning of $C_{\rm w}$, which is the most important parameter in
the present study, is clearly explained and discussed in Appendix A.
About 0.4\% of dust can be lost through radiation-driven stellar
wind within $10^6$ yr in a model with $C_{\rm w}=0.01$.

The value of $C_{\rm w}$ can be  different between small and large
C-grains and Si-grains. For example, the wind rate for small C-grains
($C_{\rm w,s, C}$; $i=$C)  is
given as:
\begin{equation}
W_{\rm d, s, C}(t)=C_{\rm w, s, C}  S(t) M_{\rm d, s, C},
\end{equation}
where $C_{\rm w, s, C}$ is the value of $C_{\rm w}$  for small C-grains.
As demonstrated by recent numerical simulations of dust wind in 
galaxies (Bekki 2015), it depends on a number of factors whether 
dust can be ejected and removed from galaxies beyond their halo regions.
For example, gas-dust interaction through gaseous drag can prevent
dust from being ejected from gas disks of galaxies.
Since this gas-dust
interaction can not be modeled in the present one-zone models,
we regard $C_{\rm w}$ as a free parameter.
Thus, we investigate the effects of dust wind in the evolution
of dust by changing the $C_{\rm w}$ parameters for different grains.

In the present study, we assume that dust wind can occur only in
strong starburst phases (SB1, SB2, and SB3):
no dust can be removed from a galaxy
in the quiescent star-formation.
This assumption is quite reasonable and realistic, because
dust wind is observed in starburst galaxies like M82 
(e.g., Kaneda et al. 2010).
The strength of radiation pressure of stars in a galaxy
is assumed to be proportional to the total luminosity
of the  galaxy, and the total luminosity is calculated  as follows:
\begin{equation}
S(t) = \int^t_0
f_{\rm M/L}^{-1}(t-T) M_{\rm ns}(T) dT,
\end{equation}
where $f_{\rm M/L}$ is the mass-to-light ratio ($M/L$) of a single stellar population (SSP)
and $M_{\rm ns}(T)$ is the total mass of stars formed at $t=T$.
Since the $M/L$ of a SSP is a function of age and metallicity, we need to use a stellar population
synthesis code in order to properly calculate $f_{\rm M/L}$ at each time step.
Accordingly,  we use the code MILES  (Vazdekis et al. 2010) 
for the M/L estimation of all models
in the present
study.  We adopt the SSP table for the Salpeter IMF and [Z/H]=$-0.4$ from
the MILES and estimate $M/L$ for stars with different ages in all models.

Since the parameter $C_{\rm w}$ is already introduced in equation
(23),  the above equation (25) is slightly different from that
(with a parameter  $C_{\rm r}$) used in  BT14.
The radiation-driven  wind can remove dust from a galaxy,
and thus $C_{\rm w}$ is non-zero only for
$t_{\rm sb1,s} \le t \le t_{\rm sb1,e}$,
$t_{\rm sb2,s} \le t \le t_{\rm sb2,e}$,
and $t_{\rm sb3,s} \le t \le t_{\rm sb3,e}$: $C_{\rm w}=0$ in
the quiescent SF phase.
We have run a large number of models with different values of
$C_{\rm w, s, C}$,
$C_{\rm w, l, C}$,
$C_{\rm w, s, Si}$,
and $C_{\rm w, l, Si}$ so that we can find models that can explain
the observed extinction curves of the LMC and the SMC.
We mainly show the results of  three ``fiducial''
 models with a particular combinations
of these four parameters for the SMC, the LMC, and the MW, because
they can reproduce the observed extinction curves.

\subsection{Derivation of extinction curves}

Based on the total masses (or fractions) of small and large
C-grains and Si-grains in a model,  we can draw the extinction
curve by using the two-size approximation method by H15.
Since the details of the method to derive extinction curves
from dust masses and compositions are already given in H15,
we briefly describe it here.
We adopt the ``modified'' log-normal model for the grain size distribution
of small and large grains, and the functional form  ($n_{i,j}(a)$)
for the distribution is given as follows:
\begin{equation}
n_{i,j}(a)=\frac { C_{i,j} }{ a^4 } {\exp} \{ - \frac{ [\ln (a/a_{0,i,j}) ]^2 }
{ 2\sigma^2 }  \},
\end{equation}
where $a$ is the grain size, 
subscript $i$ describes 
the small ($i$=s) or large ($i$=l) grains, 
subscript $j$ describes 
carbonaceous ($j$=C) or silicate ($j$=Si) grains, 
$C_{i,j}$ is the normalization
constant, and $a_{0,i,j}$ and $\sigma$ are the central grain radius
and the standard deviation of the log-normal distribution, respectively.
It should be noted here that the mass distribution ($\propto a^3 n_{i,j}(a)$) 
is log-normal in the adopted model.

We adopt  $a_{0,\rm,s, C} = a_{0, \rm, s, Si} = 0.005$ $\mu$m,
$a_{0,\rm,l, C} = a_{0, \rm, l, Si} = 0.1$ $\mu$m,
and $\sigma=0.75$ so that the size distribution function
can roughly cover the small and large grain size ranges.
The normalizing constant for small and large C-grains and Si-grains
are determined by the following equation:
\begin{equation}
\mu {\rm m}_{\rm H} D_{i,j} 
= \int_0^{\infty} \frac{4}{3} \pi a^3 sn_{i,j}(a) da,
\end{equation}
where $\mu=1.4$ is the gas mass per hydrogen nucleus, 
$m_{\rm H}$ is the hydrogen atom mass,
and $s$ is the material density of a  dust grain.
Accordingly, we can derive $C_{i,j}$ for $D_{i,j}$ derived from
one-zone chemical evolution models by using the above equation.

Based on $n_{i,j}$ estimated from $D_{i,j}$ at each time step in
a chemical evolution model,
we can derive the dust extinction curve at the time step.
The extinction  at wavelength $\lambda$ in units of magnitude
($A_{\lambda}$) normalized to the column density of hydrogen
nuclei ($N_{\rm H}$) is described as follows (H15):
\begin{equation}
\frac{ A_{\lambda} }{ N_{\rm H} } =  
 2.5 \log {\rm e} \sum_{i} \sum_{j}
\int_0^{\infty} n_{i,j} \pi a^2 Q_{\rm ext}(a,\lambda),
\end{equation}
where $Q_{\rm ext}(a,\lambda)$ is the extinction efficiency factor.
The values of this $Q_{\rm ext}$ factor at each wavelength
and for each grain are evaluated by using the Mie theory
(Bohren \& Huffman 1983) and the same optical constants for silicate
and carbonaceous dust in Weingartner \& Draine (2001).
We adopt $s=3.5$ g cm$^{-3}$ for small and large Si-grains
and $s=2.24$ g cm$^{-3}$ for small and large C-grains.
As shown in H15,  this extinction model can reproduce
reasonably well the observed MW extinction curve.

\subsection{The model parameters for the MCs and the MW}

We first investigated many models with different model parameters
for dust growth, shattering, and coagulation and found that no models
can reproduce  the observed extinction curves 
without  the 2175 \AA  $\;$  bump  (SMC) and the very weak one (LMC).
We then considered the effects of dust wind on the evolution of
dust extinction curves by incorporating the dust ejection/removal
process associated with dust wind into the one-zone models. 
Hou \& Hirashita (2015) have adopted one-zone models without dust wind
for deriving dust extinction curves of galaxies
and found that the models with a wide range of 
parameters for dust-related physical processes  (e.g., coagulation)
can not reproduce the observed extinction
curves without the 2175 \AA  $\;$  bump.
Therefore, it seems that standard chemical evolution models without
dust wind can not reproduce the extinction curves of the MCs.

After extensive investigation of models with dust-related physical
processes (e.g., selective destruction of C-grains by SNe and dust wind),
we could finally find that  models with dust wind  can reproduce better
the observed
extinction curves of the MCs.
Therefore, we focus on the results of the models
with  dust wind in the present study.
Although it is possible that physical processes other than dust wind
can be responsible for the origin of the extinction curves of the MCs,
we focus exclusively on the effects of dust wind on
the evolution of extinction curves of the MCs.

We consider that the efficient ejection/removal of dust grains
from the SMC and the LMC can be the key physical process for the
origin of their dust extinction curves. 
We therefore (i) change the parameters of dust wind
and (ii) adopt reasonable values 
for other model parameters (e.g., for dust growth)
so that we can more clearly elucidate the key factor for the presence
or absence of the 2175 \AA  $\;$  bump.
The model parameters for  star formation (e.g., burst epoch) and
dust-related physical processes (e.g., dust accretion
timescale)  are summarized in Table 1 and Table 2, respectively,
for the MCs and the MW.
BT12 have already discussed  how the chemical evolution (thus dust evolution)
histories of the MCs depend on star formation parameters,
and H15 
has discussed how the dust evolution of a galaxy depends on the model parameters
for  dust growth, destruction, shattering and coagulation.
Furthermore, we have confirmed that only some models with dust wind 
can explain the observed extinction curves of the MCs.
Therefore, we focus exclusively on how dust wind plays  a key role in controlling
the extinction curves of the MCs.

In determining the three burst epochs, we consider both recent observational
and theoretical results on the star formation and chemical evolution
histories of the MCs. 
Harris \& Zaritsky (2004) investigated the ages of stars in the SMC based
on UBVI photometry from their Magellanic Clouds Photometric Survey and
found significant rises in the mean star formation rate around 0.06,
0.4, and 2.5 Gyr ago. Using gas dynamical simulations of the SMC
interacting with the LMC and the MW, Yoshizawa \& Noguchi (2003) predicted 
that the SMC experienced starburst events about 0.2 Gyr and 1.5 Gyr ago.
Rubele  et al. (2015) have recently detected  the starburst populations
with ages of 1.5 and 5 Gyr in the deep images of the SMC from the VISTA
survey of the Magellanic Clouds in the YJK$_{\rm s}$) bands.
Guided by these observations, we assume that the SMC experienced
three enhanced star formation episodes at $t_{\rm sb, 1, s}=8.5$, 11.5, and 12.85 Gyr
(assuming the current age of 13 Gyr for the LMC).

The LMC also appears to have experienced at least three  major epochs of
significantly enhanced star formation. 
Rubele et al. (2012) investigated the AMR of the LMC stellar populations
and found that the SFR peak around 2 Gyr ago, though the AMR appears to be
different between different local regions.
Based on the results of  the gas dynamical simulations of the LMC
over the past 6 Gyr,
Bekki et al. (2004) suggested that 
a starburst, which can form globular clusters,
could have occurred in the LMC about 3 Gyr ago  owing to the LMC-SMC-Galaxy
tidal interaction.
Harris \& Zaritsky (2009) found very strong enhancement of mean
star formation from $10^8$ yr ago  to the present in the blue arm,
Constellation III, and 30 Doradus region of the LMC.
Thus, we assume that the LMC experienced 
three enhanced star formation at $t_{\rm sb, 1, s}=10.0$, 11.5, and 12.85 Gyr
(corresponding to 3, 1.5, and 0.15 Gyr ago, respectively).

Table 3 summarizes the  model parameters of dust wind in the three
galaxies. Although we investigated much more models than those 
listed in this table,
we show only these representative ones, because they grasp some
essential ingredients of the dust wind effects on the extinction curves
of galaxies. We mainly describe the time evolution
of the extinction curve in the fiducial model
for the SMC (S1). This fiducial model is chosen, because it can
best reproduce the observed extinction curve of the SMC.
As a comparative experiment, the model without dust wind
(yet with starburst; S2) is investigated for the SMC. These models with
and without dust wind enable us to demonstrate more clearly how
the dust wind can change the shapes of the extinction curves of the SMC.
The models with different $C_{\rm w, s, C}$ (S3, S4, and S5) are investigated
so that the importance of ejection/removal of small C-grains in the evolution
of the extinction curve of the SMC can be more clearly described.
The models without starbursts (S6$-$S9) enables us to understand the importance
of starburst dust wind in the evolution of the extinction curve.

We show the results of 
the LMC models with and without dust wind also in order  to 
clarify the physical reasons for the observed weak 2175 \AA  $\;$ extinction
bump in the LMC.
We show the results of the MW model (M1)
without starburst (thus without dust wind) only,
because there is little observational 
evidence for recent strong starburst in the MW.
This MW model and those for the MCs (S1 and M1) 
can combine to demonstrate that the present models can explain 
not only the extinction curve of the MW but also those of the MCs.
Figure 1 shows the time evolution of $M_{\rm g}$ (normalized), $f_{\rm g}$
(gas mass fraction),
and SFR (normalized) for the fiducial SMC, LMC, and MW models.


\section{Results}

\subsection{SMC}
Figures 2 and 3 show the time evolution of the extinction curve 
and the relative fractions of dust grains,  respectively,
in the standard SMC
model (S1) in which the SMC can experience three strong starburst (SB) events.
Before the first SB event (SB1), the star formation rate of the SMC can be kept
low so that chemical evolution can proceed very slowly ($T<8.5$ Gyr).
In this low-metallicity phase (pre-SB1 phase), 
stellar dust production dominates the dust content 
(i.e., yet little dust  growth) since
the dust abundance is too low for other dust processing
mechanisms (in particular, shattering) to work efficiently.
As a result of this,  the ISM dust of the SMC is dominated by
large grains. The SMC thus can show a rather flat extinction curve
without the  2175 \AA  $\;$  bump in this pre-SB1 phase.

As the gas-phase metallicity becomes higher 
through more rapid chemical evolution
during SB1,  the mass fraction of small grains becomes larger  owing to
the interplay between shattering and accretion  ($T=9.8$ Gyr).
The extinction curve can therefore show a slightly steeper rise in the FUV regime
and the 2175 \AA  $\;$ bump can be appreciably seen at this post-SB1 phase.
The extinction curve becomes progressively steeper and the 2175 \AA  $\;$ bump become
more prominent after the SB1 ($T=11.5$ Gyr). 
The fraction of cold gas can become higher at this epoch,
because $D$, which can determine the gas fraction is higher.
The fraction of small Si-grains
can be as large as that of large Si-grains at this pre-SB2 phase, 
because the higher metallicity and the higher cold gas fraction combine
to allow the more rapid growth of the small grain abundance.
These behaviors in the evolution of Si grains are 
apparently similar to those reported
in Asano et al. (2014).

Although the fraction
of small Si-grains can be temporarily lower owing to the efficient destruction
of dust by SNe during the second starburst phase (SB2),
it can  start to rise again soon after SB2 ($T=12.5$ Gyr).
The total mass of  small Si-grains becomes 1.6 times
larger than that of large Si-grains at $T=12.5$ Gyr so that 
the mass fraction of small Si-grains can be the largest in the SMC history.
Consequently, the extinction curve of the SMC in this model
at $T=12.5$ Gyr shows 
a very steep rise that is even steeper than the
observed  SMC  extinction curve  and thus inconsistent
with the observed curve.
Although small C-grains can also suffer a significant loss of their mass owing to the grain
destruction by SNe during SB2, their mass can steadily and rapidly 
increase through the accretion of
gas-phase metals after SB2. This increase  of small C-grain abundance is responsible
for the strong 2175 \AA  $\;$ bump in the extinction curve at $T=12.5$ Gyr.
  
The fraction of small Si-grains can become smaller owing to their destruction
by SNe and ejection from the SMC through radiation-driven dust wind 
in the third starburst phase (SB3) so that
the extinction curve at $T=12.9$ Gyr can be slightly less steeper than
that at $T=12.5$ Gyr. The dust wind effects are more significant in this 
SB3, mainly because the total luminosity of stars in the SMC  takes its maximum.
Although a significant fraction of small C-grains can be removed from
the SMC through dust wind,  the extinction curve still clearly shows 
the 2175 \AA  $\;$ bump in the early phase of this SB3.  The 2175 
\AA  $\;$ bump can completely
disappear only after most of the small C-grains can be removed from 
the SMC through dust wind at $T=12.9$ Gyr. The final extinction curve
at $T=13$ Gyr is very similar to the observed one both in the lack of
the 2175 \AA  $\;$ bump and in the steep linear rise at FUV wavelength. Thus the present
standard SMC model with dust wind can reproduce quite well the observed
extinction curve.

As clearly shown in Figure 3,  the time evolution of C-grains is  
different between the SMC models with and without dust wind (S1 and S2,
respectively) after SB2: the fractions of small and large C-grains
and small Si-grains (relative to large Si-grains) are not
so different between the two models during and after SB1 and SB2. 
This suggests that  the SMC can not show 
the extinction curve without the 2175  \AA  $\;$ bump
until quite recently. This also suggests (i) that star-bursting dwarf galaxies
do not necessarily show extinction curves without the 2175 \AA  $\;$ bump
and (ii) that dust wind needs to be strong enough to remove most (90 percent)
of small C-grains
in star-bursting dwarfs with no 2175 \AA  $\;$ bump in their extinction curves.
It is thus highly likely that starburst dwarfs have different strength
of the 2175 \AA  $\;$ bump.

Figure 4 shows that the final extinction curve
in the  SMC model with three SB events yet no dust wind
has the conspicuous 2175 \AA  $\;$ bump.
The SMC models with lower dust wind efficiency ($C_{\rm w, s, C}$=0.01 and 0.03 
for S4 and S5, respectively)
also show the 2175 \AA  $\;$ bump and the FUV rise that is less consistent with
the observed extinction curve of the SMC.
The SMC model with higher dust wind efficiency 
($C_{\rm w, s, C}$=0.1, S6) can reproduce
the observed extinction curve of the SMC as well as the standard SMC model.
These results confirm  that the removal
of small C-grains from the SMC through dust wind is a viable mechanism for the observed lack
of the 2175 \AA  $\;$  bump in the extinction curve of the SMC.

The model with no starburst thus no dust wind (S7) shows the
conspicuous 2175 \AA  $\;$ bump
and a steeper FUV rise,
which clearly demonstrates that dust wind can significantly influence the shape
of the extinction curve of the SMC. 
As shown in the models S8 and S9,
the presence  or absence of the 2175 \AA  $\;$ bump 
dose not depend on whether SB1 or SB2 occurs in the SMC.
However, the 2175 \AA  $\;$ bump of the extinction curve
is too strong in the model S10 without SB3,  
which suggests that the strength of the 2175 \AA  $\;$ bump can depend on
whether the SMC experiences SB3 with dust wind.
These results thus confirm that the last SB with strong dust wind
about 0.2 Gyr ago is a key physical process
for the observed characteristic extinction curve of the SMC.
The results of the SMC models with different strengths of starbursts
are discussed in Appendix B.
The metallicity evolution of the SMC (LMC and MW) in the present new
models with dust evolution is discussed briefly in
Appendix C.

\subsection{LMC}

Figure 5 presents the time evolution of the extinction curve of the LMC
in the model L1 with three SB events.  Gas can be more rapidly converted
into new stars in this L1 than in the SMC models 
(owing to the adopted higher SF efficiency) so that
chemical abundances can be higher even well before the first SB event ($T=8.4$ Gyr).
As shown in Figure 6,   small C- and Si-grains can start to grow earlier 
in this L1, and their mass fraction can take their peak values  before SB1.
Consequently,  the extinction curve can have a rather conspicuous 2175 
\AA  $\;$ bump
and a very steep FUV rise at $T=8.4$ and 9.8 Gyr.
Since SNe can efficiently destroy dust grains during SB1 owing
to high SF rates (i.e. high SN rates), the 2175 \AA  $\;$ bump
becomes weakened  after SB1 ($T=11.5$ Gyr). However, such destruction of
small grains can not severely weaken the 2175 \AA  $\;$ 
bump after SB1 and even after
SB2 ($T=12.5$ Gyr).

The 2175 \AA  $\;$ bump of the LMC becomes less conspicuous during SB3
owing to the combination effect of dust destruction and radiation-driven
dust wind ($T=12.9$ Gyr).   The shape of the extinction curve of the LMC can not become
quite similar to the observed one until the end of SB3 ($T=13$ Gyr).
The rather weak 2175 \AA  $\;$
 bump at $T=13$ Gyr in this model is due to the more 
preferential removal of small C-grains from the LMC through 
radiation-driven dust wind during the longer starburst (SB3).
The major difference in the dust abundances between
this LMC model L1 and the SMC model S1 is that
the final fractions
of small and large C-grains are both slightly larger in L1 than in S1
(See Figures 4 and 6).
These differences can cause a difference
in the extinction curves between the LMC and the SMC in the present study.

Figure 7 compares between the extinction curve of L1 with dust wind
and that of L2 without dust wind. Clearly, L1 with dust wind
can better reproduce the observed extinction curve of the LMC,
which confirms that the efficient removal
of C-grains through radiation driven dust wind is essential for
explaining the origin of the extinction curve of the LMC.
Previous numerical simulations of the MCs (e.g., Yoshizawa \& Noguchi 2003;
Bekki \& Chiba 2005) suggested that the observed enhanced star formation
rates in the MCs (e.g., Harris \& Zaritsky 2006, 2009) can be due
largely to the LMC-SMC-Galaxy tidal interaction in the past 3 Gyr. 
Thus the results in Figures 2-7 suggest that
the origin of the observed extinction curves of the LMC and the SMC
can be closely associated with  the past  active star formation
triggered
by tidal interaction among the three galaxies.

\subsection{MW}

It is instructive for the present study to show whether the new models
for dust evolution of galaxies can reproduce the observed extinction
curve of the MW just by changing the model parameters.
Figure 8 describes the time evolution of the MW model M1 
without starburst events (thus without dust wind).
In the very early chemical evolution history of the model M1 ($T=1$ Gyr),
the extinction curve is rather flat without the clear 2175 \AA  $\;$ bump
owing to the very small fraction of small C-grains.
The extinction curve  of the model M1 
evolves to have the 2175 \AA  $\;$ bump and a steeper rise
at $1/\lambda > 6$ $\mu$m$^{-1}$ only 2 Gyr after the commencement
of its star formation (i.e., disk formation).
The overall feature of the extinction curve becomes quite similar
to the observed one for the MW at $T=5$ Gyr, which implies that
MW-like galaxies at higher redshifts might already have extinction curves
similar to that of the present MW.

The 2175 \AA  $\;$ bump is a bit too conspicuous at $T=5$ Gyr
in comparison with the observed one, and the bump does not
evolve significantly with time after $T=5$ Gyr.
This is mainly because the relative mass fractions of small Si-grains
and small and large C-grains (with respect to large Si-grains)
do not change significantly after $T=5$ Gyr. 
The overall shape of the extinction curve at $T=13$ Gyr
becomes closer to that of the observed one.
However, the final shape of the 
2175 \AA  $\;$ bump is still slightly different from
the observe one (e.g.,  more remarkable bump),
which implies that the present models do not reproduce the
dust abundances of the MW completely well. 
It is our future study to reproduce better the observed extinction
curve of the MW by using a more sophisticated dust evolution model.

These results imply that luminous disk galaxies like the MW
are likely to show clear 2175 \AA  $\;$ bumps in their extinction
curves, because a significant amount of dust is unlikely to be removed
from the galaxies owing to their deep gravitational potential. 
Although a small  amount of dust and metals could be removed/ejected
from the galaxies through SN and dust wind,  such a less significant removal/ejection
process can not influence the shapes of the extinction
curves to a large extent. The present study therefore predicts that the shapes of extinction
curves depend on the masses of galaxies in such a way that more massive
galaxies are likely to have the 2175 \AA  $\;$ bump. 
It is an interesting observational question how the strength
of the  2175 \AA  $\;$ bump in a massive galaxy depends on
its physical parameter (e.g., total mass and luminosity).
It is worth noting here that Asano et al. (2014) and Nozawa et al. (2015)
attempted to reproduce the 2175 \AA  $\;$  bump of the MW 
by freely changing the mass fractions of various ISM phases.

Finally, Figure 9 summarizes the time evolution of
mass fractions of the key three dust grains with respect to large Si-grains
for the MCs and the MW.
The differences in the evolution between the three galaxies can be clearly
seen, which reflects on the time evolution of the extinction curves of
the three.
Star formation histories with or without starbursts and capabilities
to retain dust within galaxies are key determinant for the time evolution
of the extinction curves of galaxies.
Since the present study has investigated only three different 
types of galaxies (i.e., SMC, LMC, and MW),
we will need to investigate models with different star formation 
and dust evolution histories
to understand fully the origin of various extinction curves observed
in galaxies with different masses and Hubble types.

\section{Discussion}

\subsection{The origin of the 2175 \AA $\;$ bump}

We have, for the first time, demonstrated that the observed extinction
curve without  the   2175 \AA $\;$ bump
in the SMC can be reproduced, if the small C-grains can be more preferentially
lost through radiation-driven wind during the last starburst around
0.2 Gyr ago. In this ``lost small C-grain" scenario, 
the LMC can have a weak  2175 \AA $\;$ bump in its dust extinction curve,
because it can retain a larger amount of small C-grains
owing to its deeper gravitational potential in comparison with
the SMC. 
This ``lost small C-grain" scenario 
predicts that more  massive galaxies (like the MW)  are likely to 
show more distinctive  2175 \AA $\;$ bumps in their  dust extinction curves.
It should be stressed, however, that the present low-mass galaxies can show
such 2175 \AA $\;$ bumps if they have not experienced recent 
strong starbursts 
and the resultant loss of small C-grains.

The key physical process required for
the lack of  the  2175 \AA $\;$ bump in a galaxy is the more
efficient removal of C-grains than Si-grains from the galaxy.
The removal (or stripping) process of dust grains in galaxies should be
rather complicated, because gravitational dynamics, gas-dust drag,
and radiation pressure of stars can be all involved in the removal process.
Therefore, it would be a formidable task to estimate accurately the dust removal
efficiency in a galaxy for the physical parameters of the galaxy.
Thus we  investigate whether and in what physical conditions
the above two requirements can be met in 
a galaxy by using a simplified yet reasonable model.

The first question is whether these C-grains can be more efficiently removed
from a galaxy through radiation-driven wind than Si-grains.
The key parameter for this removal process in a galaxy 
is the frequency-averaged
radiation pressure coefficient ($Q_{\rm pr}^{\ast}$),
 as demonstrated in previous theoretical models  and recent numerical
simulations of dust wind in galaxies  (e.g., Ferrara et al. 1991; 
Bianch \& Ferrara 2005; Bekki 2015).
We have therefore investigated $Q_{\rm pr}^{\ast}$
for C and Si grains  by using the publicly available data
for optical properties of dust (Draine and Lee 1984; Laor \& Draine 1993) and the stellar 
population synthesis code (``MILES") developed by Vazdekis et al. (2010).
The radiation pressure coefficient
($Q_{\rm pr}$) for a grain with a radius is given as follows:
\begin{equation}
Q_{\rm pr}=Q_{\rm abs}+(1-g(\theta))Q_{\rm sca},
\end{equation}
where $Q_{\rm abs}$ and $Q_{\rm sca}$ are the absorption and scattering coefficient,
respectively, and $g(\theta)$ is the scattering asymmetry parameter. 

Using $Q_{\rm pr}$ and the flux of a SED ($f(\lambda)$),
we can estimate  $Q_{\rm pr}^{\ast}$ as follows:
\begin{equation}
Q_{\rm pr}^{\ast}=\int_{\lambda_{\rm min}}^{\lambda_{\rm max}} Q_{\rm pr} f(\lambda) 
d\lambda,
\end{equation}
where $\lambda_{\rm min}$ and  $\lambda_{\rm max}$  are set to be 3340.5  \AA  $\;$  
and 7409.6 \AA, respectively, in MILES.
The above wavelength range is publicly available one in MILES and $f(\lambda)$
is the normalized flux 
(i.e., $\int_{\lambda_{\rm min}}^{\lambda_{\rm max}} f(\lambda)=1$).
Figure 10 shows the wavelength-dependence of radiation pressure coefficient
($Q_{\rm pr}$) for silicate and graphite with the radii of 0.01 $\mu$m and 0.1$\mu$m
and the SED of a galaxy for a metallicity ([Z/H]) of $-0.7$ dex and 
an age of 63 Myr (which is the youngest stellar population that can
be adopted in MILES).  We use this SED for a single stellar population (SSP) to
mimic the SMC with young starburst components.
We discuss just this model with two representative dust radii as an example,  though, ideally speaking,
we should show the results for all dust radii.

Clearly, $Q_{\rm pr}^{\ast}$ is higher in graphite than in silicate
for the adopted SED and dust radii (0.01 $\mu$m and 0.1 $\mu$m),
which means that graphite is more likely
to be influenced by radiation pressure of stars and thus removed
from a galaxy if the gravitational potential is sufficiently shallow.
Furthermore, the difference in $Q_{\rm pr}^{\ast}$ is significantly larger
in the small grains ($a=0.01$ $\mu$m) than in the large ones
($a=0.1$ $\mu$m), which implies that small C-grains can be much more
efficiently removed than small Si-grains.
Although the real SED of a SMC-type galaxy might be different from
the adopted SED in this investigation,
it is unlikely that  this result changes qualitatively if a more realistic
SED is adopted, given $Q_{\rm pr}$ of graphite 
always higher than those of silicate
for the relevant wavelength range.
These results thus suggest that graphite is more likely to be lost
from a SMC-type galaxy than silicate, though a full investigation
on this dust removal process for different dust grains and more
realistic galaxy spectra needs to be done in our future studies.

If a rather small mass fraction of small C-grains is responsible for
the observed lack of the 2175 \AA  $\;$ bump in the SMC,
then is there any observational evidence for that ?
Using ISOPHOTO and HiRes IRAS data for dust emission and ATCA data for HI
column gas density,
Bot et al. (2004) investigated the gas-to-dust-ratios of 
polycyclic aromatic hydrocarbon (PAH) molecules or small
grains (VSGs) for different regions
of the SMC.  
They found that the derived difference in the gas-to-dust-ratio 
of PAHs and very small grains
between the Galaxy and the SMC is three times larger than the metallicity
difference between the two galaxies. This observational result implies that
the dust depletion levels  for VSG and PAHs are low in the SMC for some
physical reasons (e.g., destruction by SNe; Bot et al. 2004). 
Since this observational study can not distinguish small C-grains from small
Si-grains, it is not clear whether the observed low depletion level
of VSGs implies that the mass fraction of small C-grains is significantly low owing
to efficient removal or destruction of such grains.
It appears to be difficult for photometric study of the SMC alone to
estimate the contributions of small C- and Si-grains to the observed
dusty SED separately.

The present study predicts that the star-forming regions of the SMC for which
there is no 2175 \AA  $\;$  bump can be dominated by Si-grains (i.e., small
fractions of C-grains). However, Welty et al. (2001) found that the 
Si appears to be {\it undepleted} in  a local region of the SMC
toward the SMC star Sk 155, which means that the local region should contain
little Si dust. If this is not just for the Sk 155 region but for other
star-forming regions generally in the SMC, 
then the present scenario and other dust models for the SMC
like Pei (1992) appears to be ruled out.
Thus it would be observationally important  to make
a robust estimation of  the dust depletion level
of Si in many star-forming regions
with and without the 2175 \AA  $\;$  bump for the SMC.

\subsection{Alternative scenarios}

A number of possible carriers of the 2175 \AA  $\;$  bump
have been extensively discussed by many authors so far (e.g., Whittet 2003).
Mishra \& Li (2015) have recently found a strong correlation
between the strength of the 2175 \AA  $\;$  bump
and the carbon depletion level ([C/H]$_{\rm dust}$) 
in the MW
and thus suggested  that the carrier of
the observed 2175 \AA  $\;$  bump is graphite or
PAH molecules.
Their results are consistent with the present ``lost small C-grains" scenario,
though they did not discuss small and large C-grains separately
in the context of the origin of the 2175 \AA  $\;$  bump.
The results imply that the time evolution of C-grains in a galaxy
can be reflected on the strength of the 2175 \AA  $\;$  bump
in the extinction curve of the galaxy.

The severely deficient small C-grain of the SMC ISM discussed
in the present study  would be just one of the promising ways
that explain the observed lack of the 2175 \AA  $\;$  bump in the SMC. 
Nozawa et al. (2015) have recently reproduced the observed
dust extinction curve without the 2175 \AA  $\;$ bump in SDSS J1048+4637
by adopting the optical constant of amorphous carbon 
for carbonaceous grains (See Figure 4 in their paper).
Although their dust extinction curve (Nozawa et al. 2015) 
is quite different from
that of the SMC,  it would be possible
that amorphous carbon dust might be responsible for the lack
of the 2175 \AA  $\;$ bump in the SMC extinction curve.
Recent observational study of dust in the LMC by Herschel (Meixner et al. 2010)
has shown that their dust model with amorphous carbon 
can better explain their observations for a more reasonable dust-to-gas-ratio.
It appears to be currently difficult to model the evolution of amorphous carbon 
by considering the possible physical effects  (e.g., radiation
fields and gaseous shock)  related to the transformation of carbon grain properties.

As demonstrated in the present study,
a way to reproduce the observed lack of the  2175 \AA  $\;$
 bump in the SMC extinction curve
is to reduce the relative contribution of small C-grains. Therefore, it does not
make much difference
 whether small C-grains can be lost either through energetic dust wind or
through some destruction process of dust grains.
If small C-grains can be more preferentially destroyed
or transformed  by SNe or 
gaseous shock
or strong radiation field in the ISM of the SMC,   
the  2175 \AA  $\;$  bump might not be observed in the SMC.
It would be possible that the last strong LMC-SMC collision/interaction 
about 0.2 Gyr ago (e.g., Diaz \& Bekki for the latest model), 
which can strongly compress the ISM of the SMC
(e.g., Bekki \& Chiba 2007), could  preferentially destroy the small C-grains.
It is our future study to investigate whether small  C-grains can be more 
preferentially destroyed by these physical processes in the ISM by using 
pc-scale numerical simulations of the ISM.

\subsection{Incorporation of four-component grain models
into numerical simulations of galaxy formation and evolution}

Interstellar dust can have significant influences on the physical processes
of galaxy formation and evolution,
such as the formation of molecular hydrogen in the ISM of galaxies 
(e.g., Gould \& Salpeter 1963) and
photo-electric heating of the ISM (e.g., Watson 1972). 
Although  these influences have been
recently included in Nbody+hydrodynamical simulations of galaxy formation
and evolution in a self-consistent manner (Bekki 2013, 2014, and 2015),
the evolution of dust sizes, which can control dust-related physical
processes of ISM (e.g., Yamasawa et al. 2011), has not been incorporated
properly into  numerical simulations of galaxy formation and evolution.
The present study has demonstrated that the evolution of dust sizes and
compositions is a key factor for the evolution of extinction curves in
galaxies. We thus suggest that if future numerical simulations of
galaxy formation and evolution can include the time evolution of
dust sizes and compositions,  such numerical studies can have the following
improvements in their predictive power.

First, such simulations with the evolution of dust sizes and compositions
can predict the SEDs of galaxies in a more consistent manner. The SEDs of dusty
galaxies are strongly influenced by the adopted extinction curves,
and the shapes of the extinction curves of galaxies depend on their star formation histories that
control their dust evolution histories.
Therefore, 
previous SED models of galaxies with a fixed extinction curve (e.g., an adoption
of the MW dust extinction curve) might be much less  self-consistent and
realistic. Construction of SEDs from age and metallicity distributions of 
stars and spatial distributions of stars in numerical simulations
has been done by many authors (e.g., Bekki et al. 1999; Bekki Shioya 2001;
Johnson 2006; Yajima et al. 2014), though they had to assume
a constant dust-to-metal-ratio and a fixed dust extinction curve. 
It is doubtlessly worthwhile for these simulations to
 include the evolution of dust sizes and compositions for a more consistent
prediction of galactic SEDs.

Second, future simulations with dust size evolution can provide more accurate
predictions of molecular hydrogen abundances in the ISM of galaxies.
As shown in sophisticated one-zone models of galaxy formation with dust size evolution
by Yamasawa et al. (2012),  the formation efficiency of molecular hydrogen
on dust grains in a forming galaxy  depends on the evolution of the dust size
distribution of the galaxy: since the molecular hydrogen formation rate on the surface
of a dust grain depends on the surface area, the net formation rate of
molecular hydrogen on dust grains in a galaxy relates to the dust extinction curve.
In order for numerical simulations to predict the dust size distribution of a galaxy,
they need to follow the time evolution of dust grains with different sizes
for each gas particle. This is a very formidable task, because such dust grain
evolution is computationally heavy even in one-zone chemical evolution models 
(e.g., Asano et al. 2014). A wise way to implement the dust size evolution
in numerical simulations of galaxy formation would be to adopt a two-size
approximation as done in this study.

Third,  the effects of photo-electric heating of dust on the thermal evolution
of ISM in galaxies can be more self-consistently investigated in numerical
simulations with dust size evolution. 
The photo-electric heating rate depends on the 
dust size distribution (not just dust abundances),
and the photo-electric heating
has been demonstrated to influence galaxy-wide  star formation through
its influences on thermal history of galactic ISM (e.g., Bekki 2015).
Therefore photo-electric heating would be one of the 
key ingredients that future numerical simulations,
in particular, those for the evolution of galaxy-wide star formation,
should include.
Thus, we suggest that incorporation of dust size evolution into numerical
simulations of galaxy formation and evolution is essential for understanding
various  aspects of galaxy formation and evolution.

\section{Conclusions}

We have constructed a new dust evolution model for the MCs by incorporating
the two-size (i.e., small and large)
approximation model for dust size distributions (H15) into one-zone
chemical evolution models with dust winds. 
The time evolution of  small and large dust grains 
has been self-consistently
investigated for each element (e.g., C, O, and Si) by considering
shattering and coagulation of these grains. 
The extinction curve of a galaxy at each time step
in a one-zone chemical evolution model has been derived from  the
relative abundances of small and large carbonaceous grains (`C-grains')
and small and large silicate grains (`Si-grains') by using
the optical constants for these grains: the galaxy
has four-component dust in the ISM.
We have searched for the model parameters
with which the observed extinction curves of the MCs can be successfully
reproduced. However, we mainly showed  the results of the models
with different wind parameters to demonstrate more clearly
the key roles of dust wind in the evolution of dust extinction curves.
The main results are as follows.

(1) The SMC extinction curve with almost  no 2175 \AA $\,$ bump
and a rather steep FUV rise can be reproduced well by the present models
with SFHs consistent with the observed ones
(`SMC model'), if the small C-grains 
can be more efficiently  ejected from the SMC through radiation-driven
dust wind during the latest starburst about 0.2 Gyr ago.
Without this selective loss of small C-grains,
the models show conspicuous 2175 \AA $\,$ feature regardless of 
model parameters for SFHs and dust physics. These results suggest
that the significantly enhanced star formation possibly triggered by the last
LMC-SMC tidal interaction can be responsible for the origin 
of the 2175 \AA $\,$ bump in the SMC.

(2) The LMC extinction curve with weak 2175 \AA $\,$ bump and 
a steeper FUV rise can be also reproduced by the present
LMC models, if the small C-grains can be ejected from the LMC more
efficiently than other dust components. The dust removal efficiency of
small C-grains required to explain the LMC extinction curve
should be by a factor of $\sim 3$ smaller than that required
to explain the SMC extinction curve.
Such a less efficient ejection of small C-grains is possible for the LMC
owing to the deeper gravitational potential well.

(3) The present model can reproduce the extinction curve
of the MW with the 2175  \AA $\,$ bump if there is no 
recent starburst event (as observed)  that can trigger strong dust wind in the model.
This is simply because small C-grains can not be preferentially
lost in the MW model with the quiescent star formation history.
We have  suggested (i) that dust grains are less likely to be ejected
from the  MW owing to its deep gravitational
potential well (in comparison with the MCs) and (ii) that  this more efficient
trapping of dust grains in the MW can be responsible for the evolution
of the extinction curve.

(4) The selective loss of small C-grains from the ISM of the MCs 
is demonstrated to be physically reasonable and realistic, 
because the frequency-averaged
radiation pressure efficiency ($Q_{\rm pr}^{\ast}$) is significantly
larger in small C-grains than in small Si-grains for
reasonable spectral energy distributions of the MCs.
Since there are other alternative mechanisms for the origin of the extinction
curves of the MCs with no/little 2175 \AA  $\;$ bump 
(e.g., selective destruction of small C-grains
and amorphous carbon grains in the MCs),
these alternative
scenarios will need to be investigated in our future papers.

(5) The present study suggests that the dust extinction curves could be
different between galaxies with different masses (i.e., difference in
the depth of gravitational potential) and
different star formation histories (e.g., with or  without starbursts), because 
the evolution of dust composition and size distributions
can be influenced by these physical properties of galaxies.
We will need a full numerical simulation including most of the dust-related
physical processes
in order to understand the dependences of the dust extinction curves
on the physical properties of galaxies.

The strength  of
the 2175 \AA  $\;$  bump, their correlations with galaxy properties,
and the influence of the bump on galactic SEDs
have been investigated for galaxies with different masses
and types at low and high redshifts (e.g., Calzetti et al. 1994; Noll et al. 2009; 
Inoue et al. 2006; Conroy 2009; Buat et al. 2011; Wild et al. 2011; Kriek \& Conroy 2013;
Mao et al. 2014; Scoville et al. 2015).
Although the present study did not discuss these observations, 
the new results based on the present  dust wind model
can provide some clues to the origin of these  observed
properties. We will discuss these issues in our  forthcoming papers.

\acknowledgments
We are  grateful to the anonymous referee for constructive and
useful comments.
HH thanks the support from the Ministry of Science and Technology (MoST) grant
102-2119-M-001-006-MY3.


\clearpage
\begin{deluxetable}{ccccccccccc}
\footnotesize  
\tablecaption{The model parameters for SF histories of the three galaxies.
\label{tbl-1}}
\tablewidth{-2pt}
\tablehead{
\colhead{  Galaxy name   } &
\colhead{  $C_{\rm q}$
  \tablenotemark{a} }   &
\colhead{  $C_{\rm sb1}$
  \tablenotemark{b} }   &
\colhead{  $C_{\rm sb2}$
  \tablenotemark{c} }   &
\colhead{  $C_{\rm sb3}$
  \tablenotemark{d} }   &
\colhead{  $t_{\rm sb1,s}$
 \tablenotemark{e} }   &
\colhead{  $t_{\rm sb1,e}$
  \tablenotemark{f} }   &
\colhead{  $t_{\rm sb2,s}$
  \tablenotemark{g} }  &
\colhead{  $t_{\rm sb3,e}$
  \tablenotemark{h} }  &
\colhead{  $t_{\rm sb3,s}$
  \tablenotemark{i} }  &
\colhead{  $t_{\rm sb3,e}$
  \tablenotemark{j} } }
\startdata
SMC & 0.005 & 0.05 & 0.05 & 0.1 &   8.5 & 8.6 & 11.5 & 11.6 & 12.85 & 12.99  \\
LMC & 0.01 &  0.05 & 0.05 & 0.1 & 10.0 & 10.1 & 11.5 & 11.6 & 12.85 & 12.99  \\
MW & 0.06 &  - & - & - & - & - & - & - & - & -  \\
\enddata
\tablenotetext{a}{The star formation coefficient ($C_{\rm sf}$)
 during the quiescent phase
(i.e., without starburst).}
\tablenotetext{b}{$C_{\rm sf}$
 during the first starburst  phase  (SB1).
The mark ``-'' in the MW model means $C_{\rm sf}$ is always $C_{\rm q}$ owing to
no starburst events.}
\tablenotetext{c}{$C_{\rm sf}$
 during the second SB phase
(SB2).}
\tablenotetext{d}{$C_{\rm sf}$
 during the third starburst phase
(SB3).}
\tablenotetext{e}{The time  (Gyr) when the first SB starts.
The mark ``-'' in the MW model means no SB events in its star formation
history.}
\tablenotetext{f}{The time  (Gyr) when the first SB ends.}
\tablenotetext{g}{The time  (Gyr) when the second SB starts.}
\tablenotetext{h}{The time  (Gyr) when the second SB ends.}
\tablenotetext{i}{The time  (Gyr) when the third SB starts.}
\tablenotetext{j}{The time  (Gyr) when the third SB ends.}
\end{deluxetable}


\clearpage
\begin{deluxetable}{cccccc}
\footnotesize  
\tablecaption{The model parameters
for the timescales for the investigated physical processes of dust.
\label{tbl-1}}
\tablewidth{-2pt}
\tablehead{
\colhead{  Physical process   } &
\colhead{  Parameter   
  \tablenotemark{a} }   &
\colhead{  small Si 
  \tablenotemark{b} }   &
\colhead{  large Si} &
\colhead{  small C} &
\colhead{  large C} }
\startdata
Accretion  & $\tau_{\rm acc, \odot}$  & 0.02 & 0.2 & 0.02 & 0.2  \\
Destruction & $\tau_{\rm SN, \odot}$  & 0.2 & 0.2 & 0.2 & 0.2  \\
Shattering  & $\tau_{\rm sh, \odot}$ & - & 0.1 & - & 0.1  \\
Coagulation  & $\tau_{\rm co, \odot}$ & 0.2 & - & 0.2 & -  \\
\enddata
\tablenotetext{a}{A simple version of the parameter description
is given (e.g., $\tau_{\rm acc}$ instead of $\tau_{\rm acc, \it i, j}$).
The value for each parameter (e.g., $\tau_{\rm acc, \odot}$) is for
the metal and dust abundances of the solar neighborhood (in the
present Galaxy).}
\tablenotetext{b}{Dust is assumed to consist
of four different types of
 grains: small and large Si-grains and small and large C-grains.
The small Si-grain is simply referred to as small Si  in this column.
Each timescale is given in units of Gyr. The mark ``-'' means
that the listed physical process is not applicable.}
\end{deluxetable}


\clearpage
\begin{deluxetable}{cccccc}
\footnotesize  
\tablecaption{The model parameters
for radiation-driven dust wind in each model.
\label{tbl-1}}
\tablewidth{-2pt}
\tablehead{
\colhead{  Model name   
  \tablenotemark{a} }   &
\colhead{  $C_{\rm w, s, Si}$  
  \tablenotemark{b} }   &
\colhead{  $C_{\rm w, l, Si}$   } &
\colhead{  $C_{\rm w, s, C}$   } &
\colhead{  $C_{\rm w, l, C}$   }  &
\colhead{  comments   } 
}
\startdata
S1  & 0.01 & 0.002  & 0.05 & 0.01 & fiducial  SMC \\
S2  & 0.0 & 0.0  & 0.0  &  0.0  & no dust wind \\
S3  & 0.01 & 0.002  & 0.01 & 0.01 &    \\
S4  & 0.01 & 0.002  & 0.03 & 0.01 &    \\
S5  & 0.01 & 0.002  & 0.1 & 0.01 &    \\
S6  & - & -  & - & - & no SB   \\
S7  & 0.01 & 0.002  & 0.05 & 0.01 & no SB1   \\
S8  & 0.01 & 0.002  & 0.05 & 0.01 & no SB2   \\
S9  & 0.01 & 0.002  & 0.05 & 0.01 & no SB3   \\
L1  & 0.01 & 0.002  & 0.03 & 0.01 & fiducial LMC  \\
L2  & 0.0 & 0.0  & 0.0 & 0.0 & no dust wind  \\
M1  & - & -  & - & -  &  fiducial MW  \\
\enddata
\tablenotetext{a}{The first capital alphabets ``S'', ``L'', ``M''
indicate SMC, LMC, and MW models, respectively.
The mark ``-'' means no dust wind owing to no starburst events
whereas the zero value (0.0) means no dust wind in spite of starburst. }
\tablenotetext{b}{The dust removal coefficient for
small and large Si-grains and small and large C-grains
is represented by
$C_{\rm w, s, Si}$,
$C_{\rm w, l, Si}$,
$C_{\rm w, s, C}$,
and $C_{\rm w, l, C}$, respectively.
}
\end{deluxetable}

\clearpage
\epsscale{0.6}
\begin{figure}
\plotone{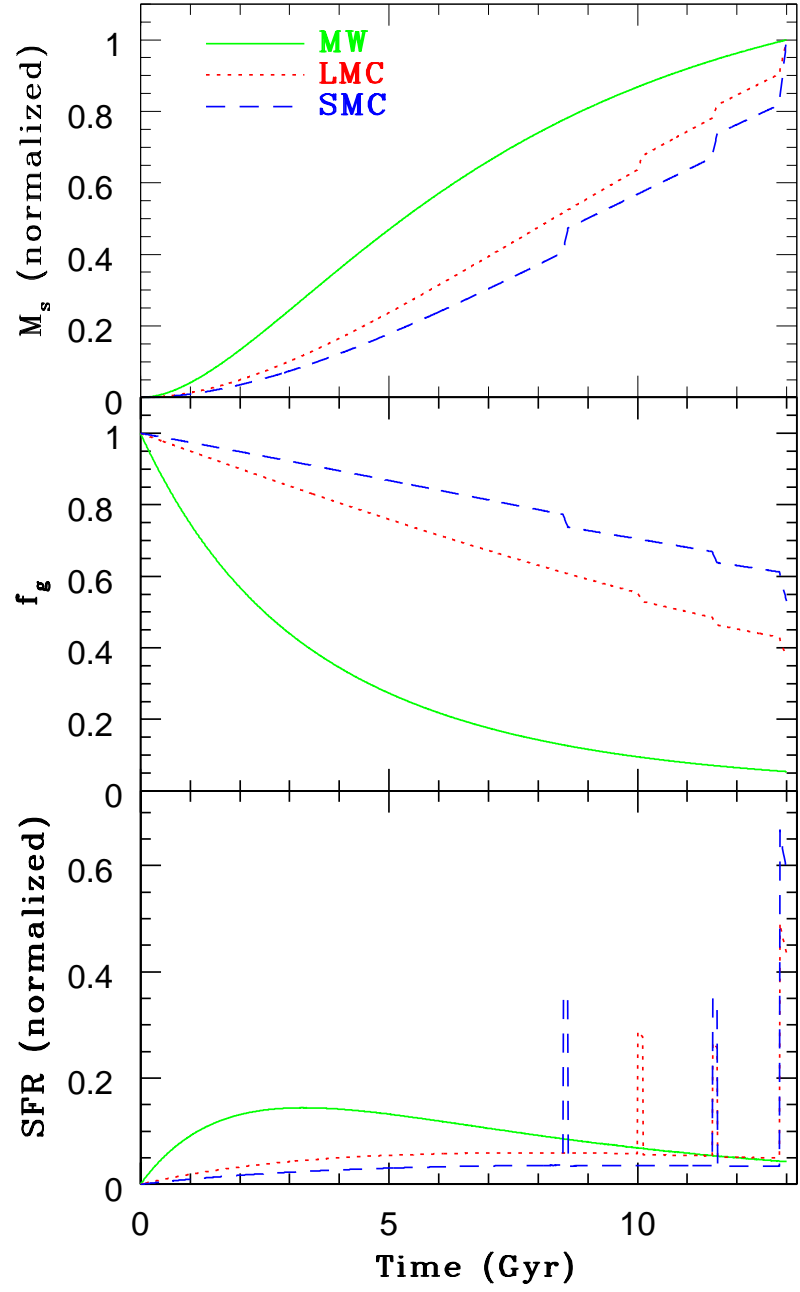}
\figcaption{
Time evolution of the total stellar mass ($M_{\rm s}$),
gas mass fraction ($f_{\rm g}$), and SFR for the MW (green solid),
the LMC (red dotted), and the SMC  (blue dashed).
The SFR at each time step is the ``normalized'' one so that
we can compare the time evolution of SFRs between the three galaxies
more clearly. 
The normalized SFR is given in units of $0.1M_{\rm unit}/t_{\rm unit}$
just for convenience,
where $M_{\rm unit}$ and $t_{\rm unit}$ are mass and time units
for a model, respectively,
and they are initial gas mass ($M_{\rm g}$) and $10^8$ yr,
respectively.
\label{fig-1}}
\end{figure}
\clearpage

\epsscale{1.0}
\begin{figure}
\plotone{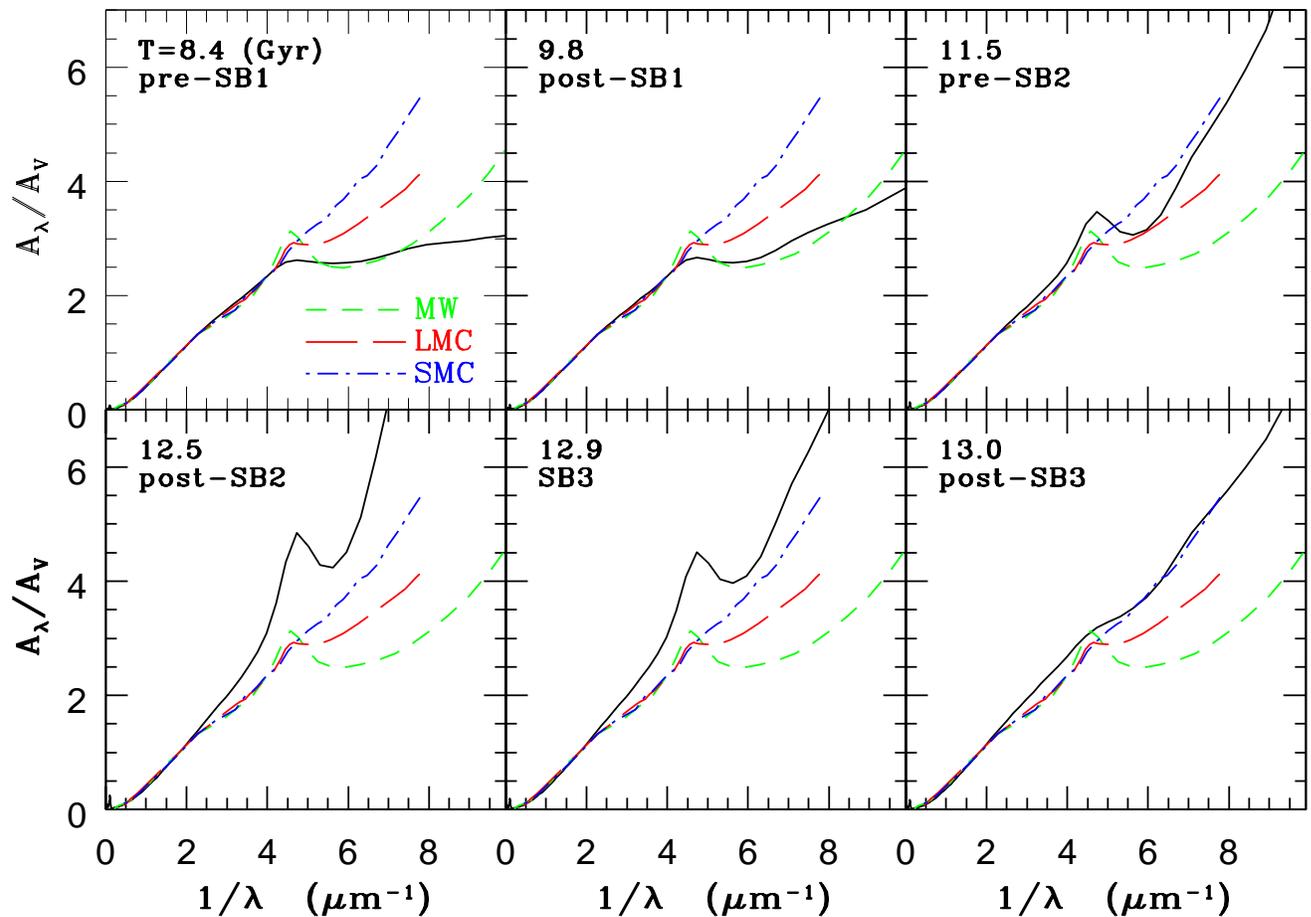}
\figcaption{
Extinction curves of the modeled SMC 
(black solid) at the selected 6 time steps,
$T=8.4$ Gyr (before the first starburst, ``pre-SB1''), 
$T=9.8$ Gyr (after the first starburst ``post-SB1''), 
$T=11.5$ Gyr (``post-SB2"), 
$T=12.5$ Gyr ( ``post-SB2''), 
$T=12.9$ Gyr (during  the third SB ``SB3''), 
and $T=13.0$ Gyr (``post-SB3'').
For comparison,  the observed extinction curves of the MW (green short-dashed),
the LMC (red long-dashed),  and the SMC (blue dot-dashed) are shown in
each frame.
\label{fig-2}}
\end{figure}
\clearpage

\epsscale{1.0}
\begin{figure}
\plotone{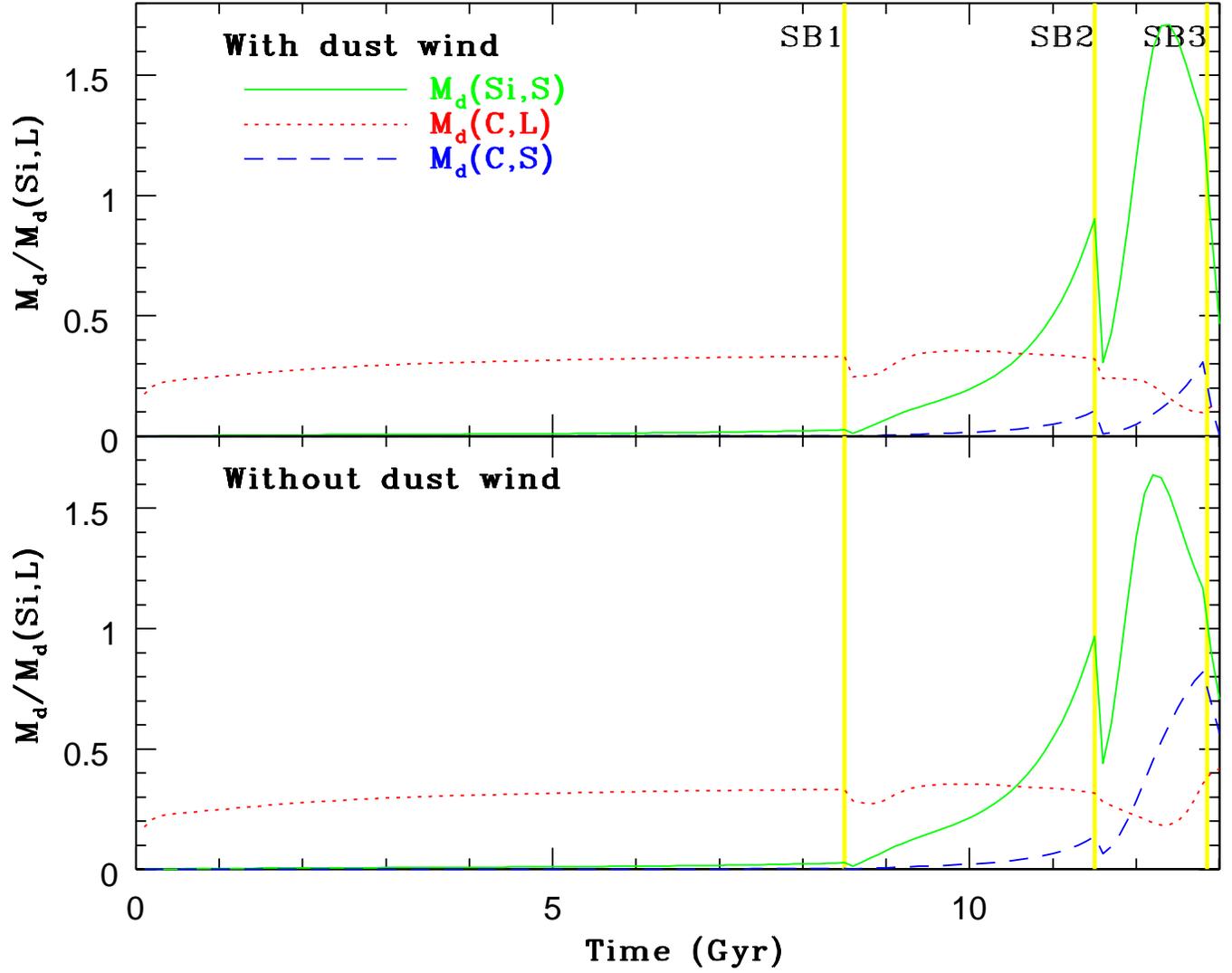}
\figcaption{
Time evolution of the total grain masses (normalized to the total mass
of large Si grains at each time step for convenience) 
for small Si-grains ($M_{\rm d} {\rm (Si, S) }$; green solid),
large C-grains ($M_{\rm d} {\rm (C, L) }$; red dotted),
and small C-grains ($M_{\rm d} {\rm (C, S) }$; blue dashed)
for the SMC model S1 with dust wind (upper) and S2 without dust wind (lower).
Three thick yellow lines indicate the epochs of starbursts
(SB1, SB2, and SB3).
\label{fig-3}}
\end{figure}
\clearpage

\epsscale{1.}
\begin{figure}
\plotone{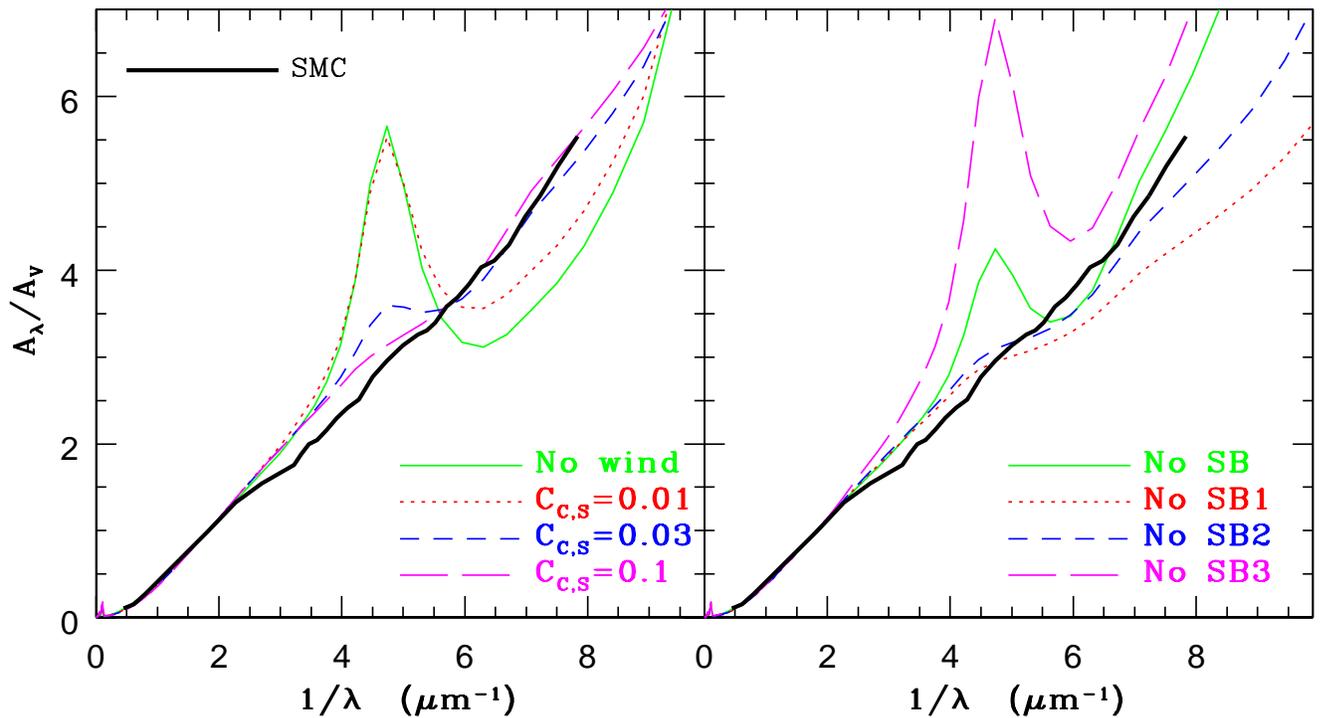}
\figcaption{
Extinction curves for the eight SMC models.
The left panel shows
S2 with starburst (SB) yet no wind (green solid),
S3 with the dust removal efficiency for small C-grains ($C_{\rm w, s, C}$)
being 0.01 (red dotted),
S4 with $C_{\rm w, s, C}=0.03$ (blue short-dashed),
S5 with $C_{\rm w, s, C}=0.1$ (magenta long-dashed).
The right panel shows
S6 with no SB thus no wind (green solid),
S7 with no SB1
(yet with SB2 and SB3; red dotted),
S8 with no SB2 (blue short-dashed),
S9 with no SB3 (magenta long-dashed).
The thick black solid line indicates the observed extinction curve of the SMC.
\label{fig-4}}
\end{figure}
\clearpage

\epsscale{1.0}
\begin{figure}
\plotone{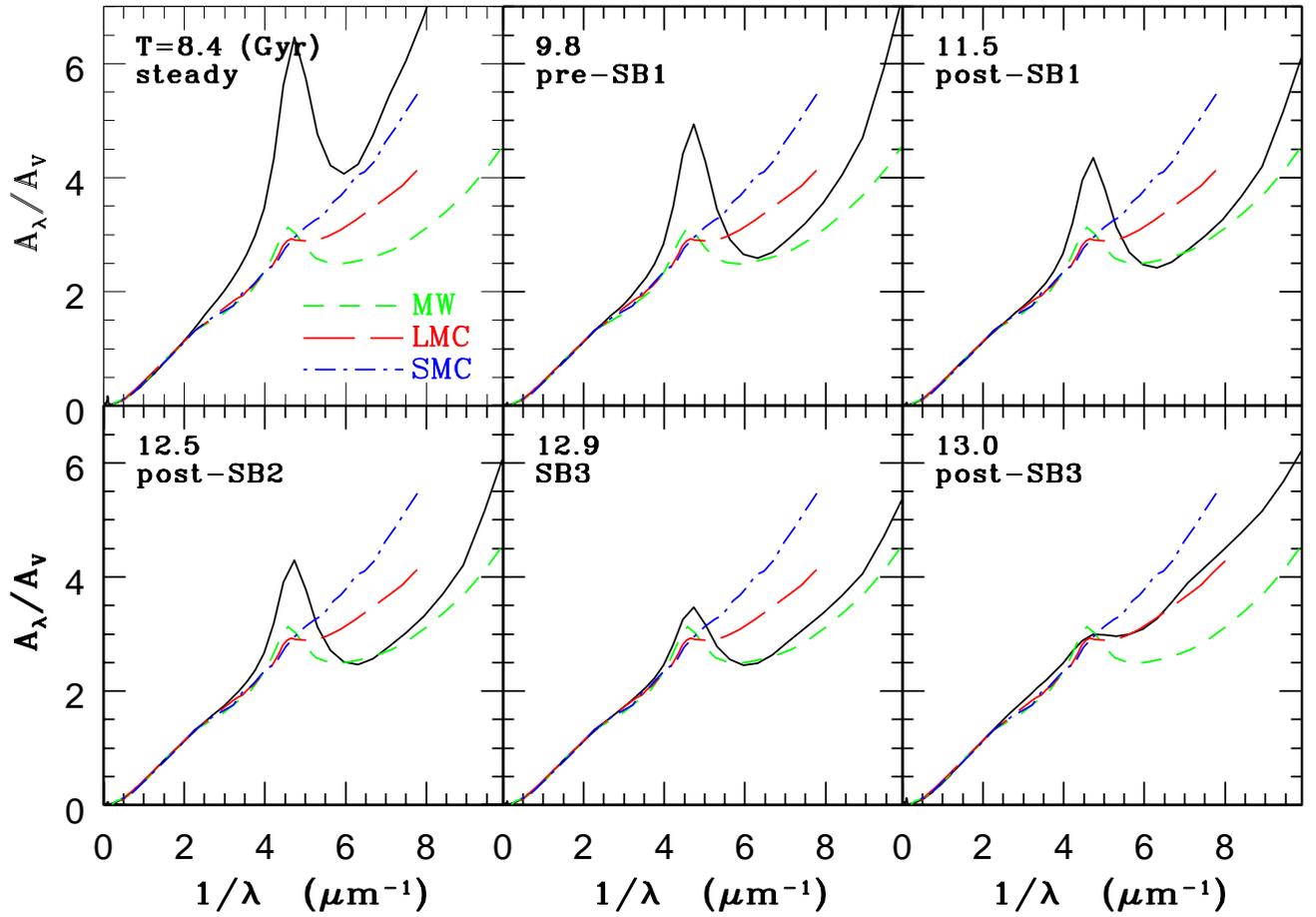}
\figcaption{
Same as Figure 2 but for the LMC model L1. The first starburst epoch
(SB1) is different between this model L1 and S1.
\label{fig-5}}
\end{figure}
\clearpage

\epsscale{1.}
\begin{figure}
\plotone{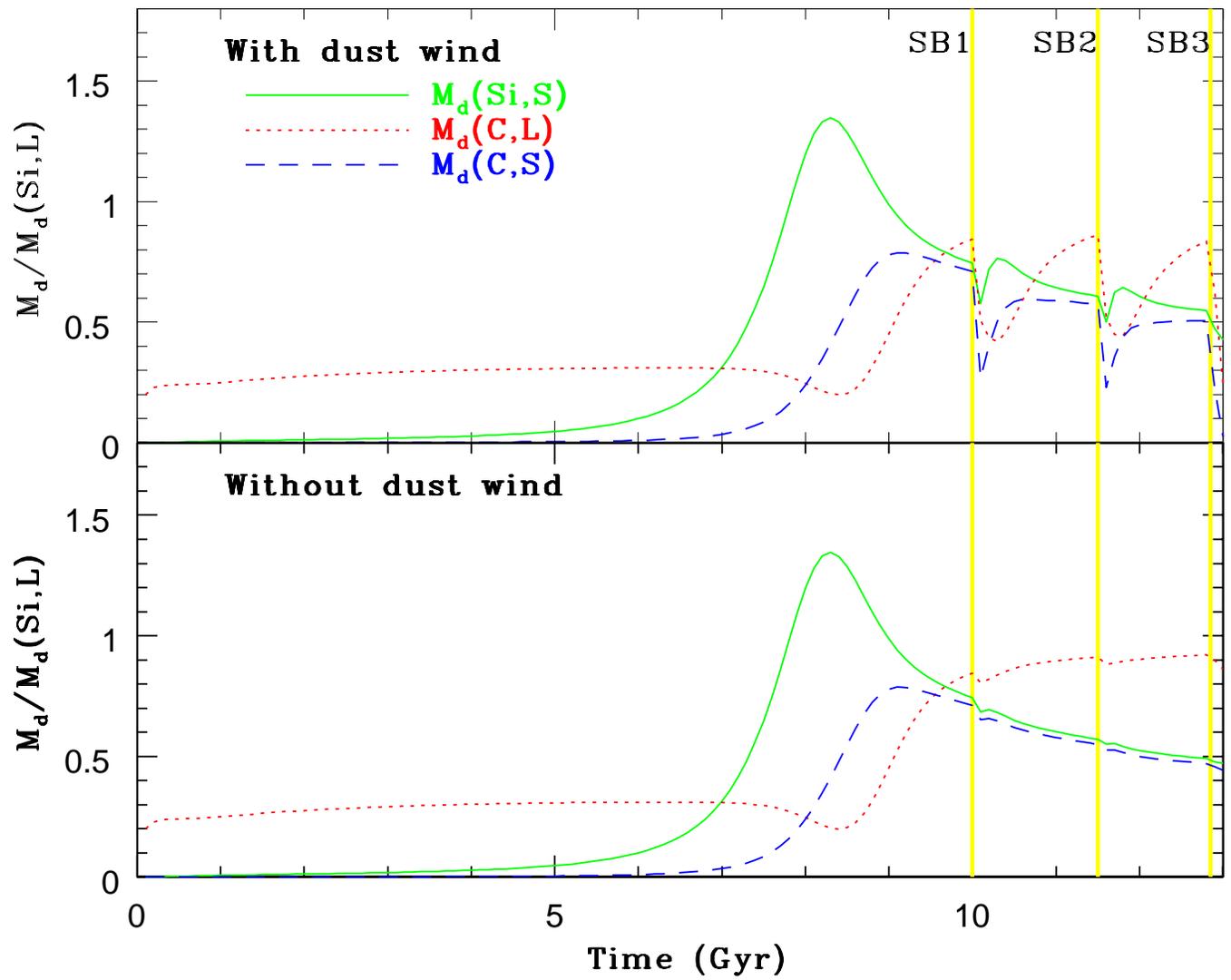}
\figcaption{
Same as Figure 3 but for the LMC models with (L1) and without (L2)
dust wind.
\label{fig-6}}
\end{figure}
\clearpage

\epsscale{1.}
\begin{figure}
\plotone{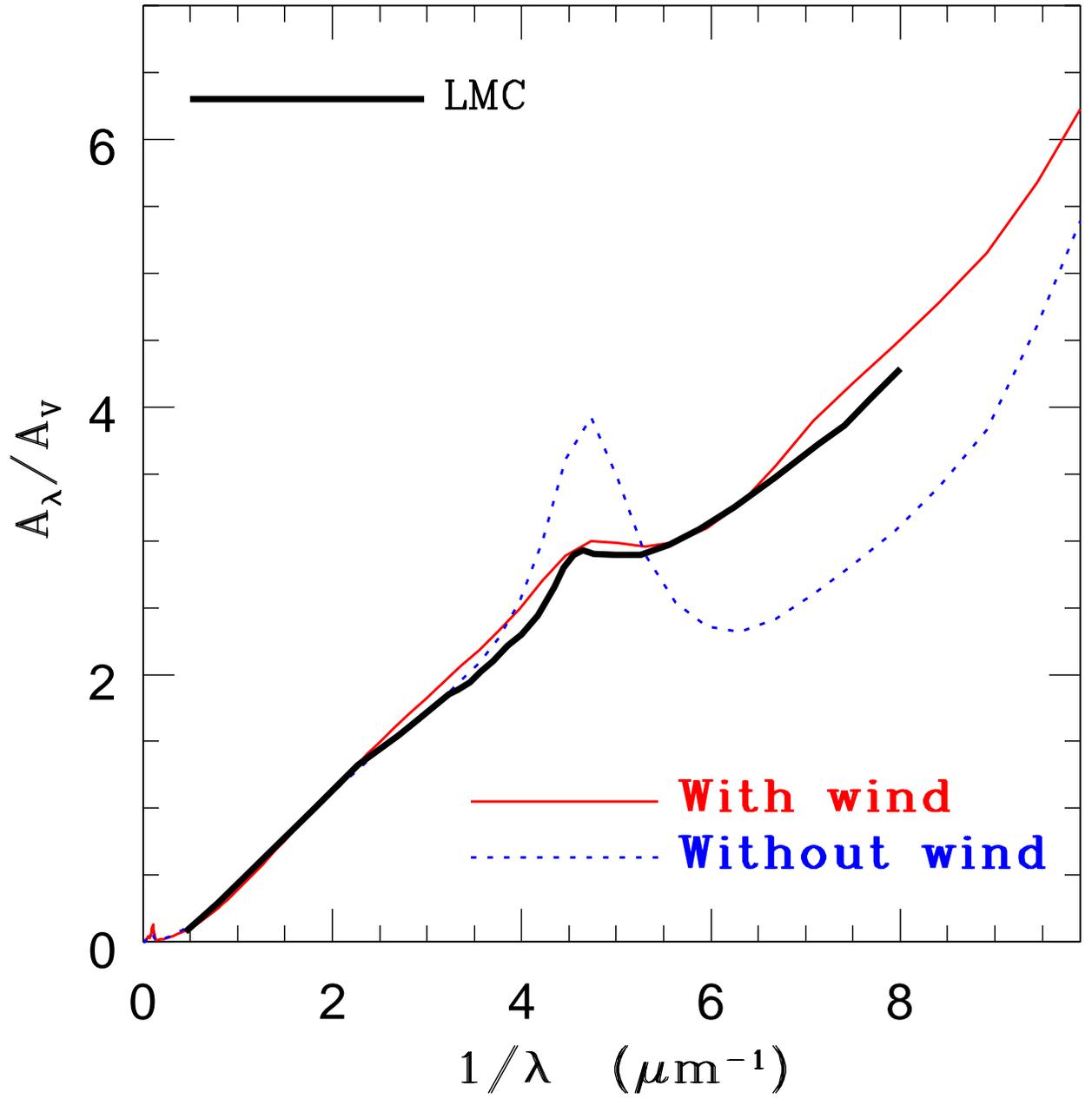}
\figcaption{
Extinction curves for the LMC models with dust wind (red solid)
and without (blue dotted) at $T=13$ Gyr (i.e, the present LMC).
For comparison, the observed extinction curve is shown.
\label{fig-7}}
\end{figure}
\clearpage

\epsscale{1.0}
\begin{figure}
\plotone{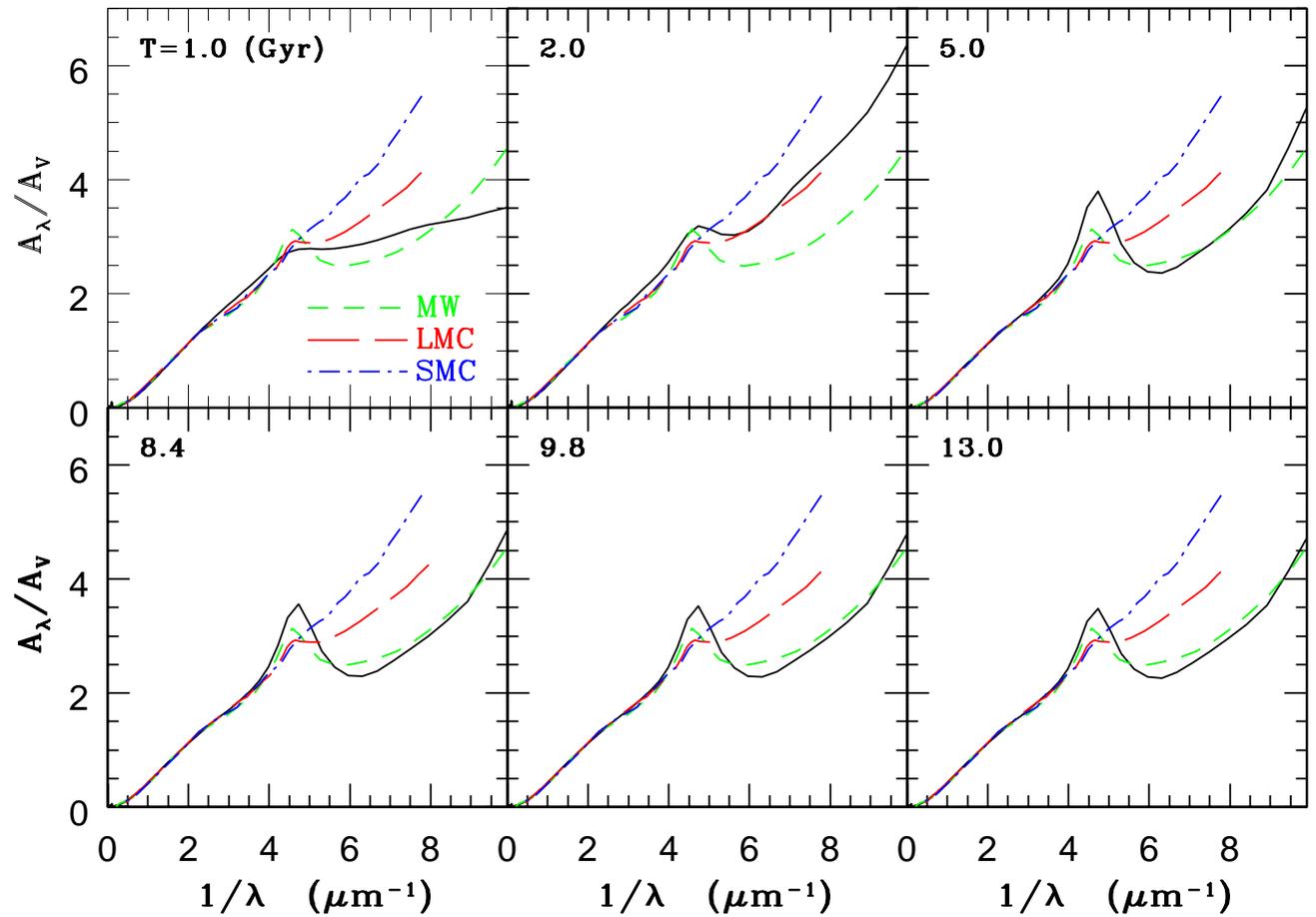}
\figcaption{
Same as Figure 2 but for the MW model M1 (without starburst and thus
no dust wind).
\label{fig-8}}
\end{figure}
\clearpage

\epsscale{.6}
\begin{figure}
\plotone{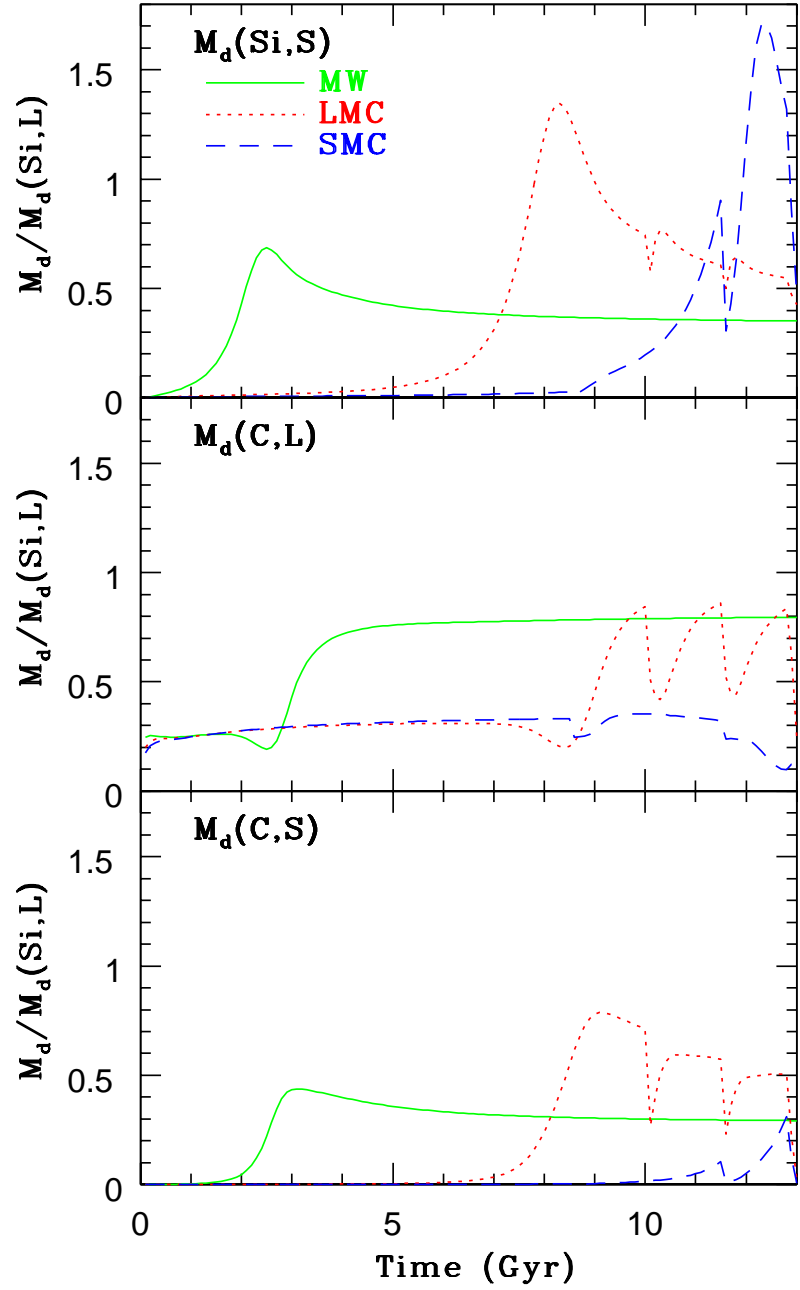}
\figcaption{
Time evolution of dust grain masses (normalized by
the mass of large Si-grain, $M_{\rm d} \rm (Si, L)$)
for small Si-grain (top),
large C-grain (middle),
and small C-grain (bottom),
for the MW M1 (green solid),
the LMC L1  (red dotted),
and the SMC S1 models (blue dashed).
\label{fig-9}}
\end{figure}

\clearpage

\epsscale{1.}
\begin{figure}
\plotone{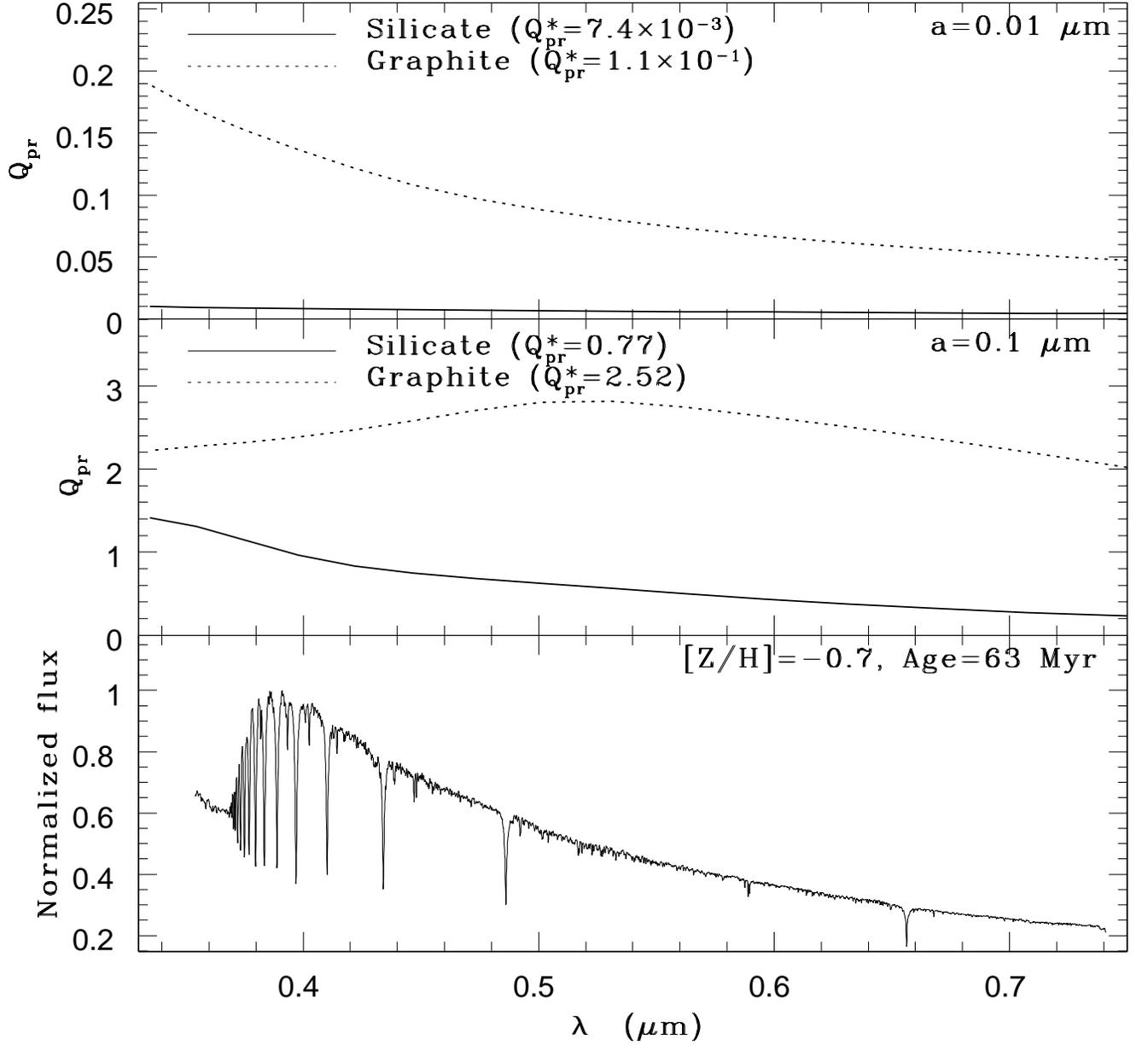}
\figcaption{
Dependence of the radiation pressure coefficient
($Q_{\rm pr}$) for silicate (solid) and graphite (dotted) with
dust radii ($a$) of 0.01 $\mu$m (top)  and 0.1 $\mu$m (middle)
on $\lambda$ and the spectral energy distribution (SED)
of a  galaxy with [Z/H]=$-0.7$ and age of 63 Myr (bottom).
This SED with the
adopted metallicity and age can mimic the SMC with a recent enhanced
star formation.
The frequency-averaged $Q_{\rm pr}$ ($Q_{\rm pr}^{\ast}$) 
for silicate and graphite are indicated in the upper part of the top and middle
panels. Clearly, $Q_{\rm pr}$ is systematically larger in graphite than
in silicate, which implies that graphite can be more strongly influenced
by radiation-driven dust wind.
\label{fig-10}}
\end{figure}

\epsscale{0.6}
\begin{figure}
\plotone{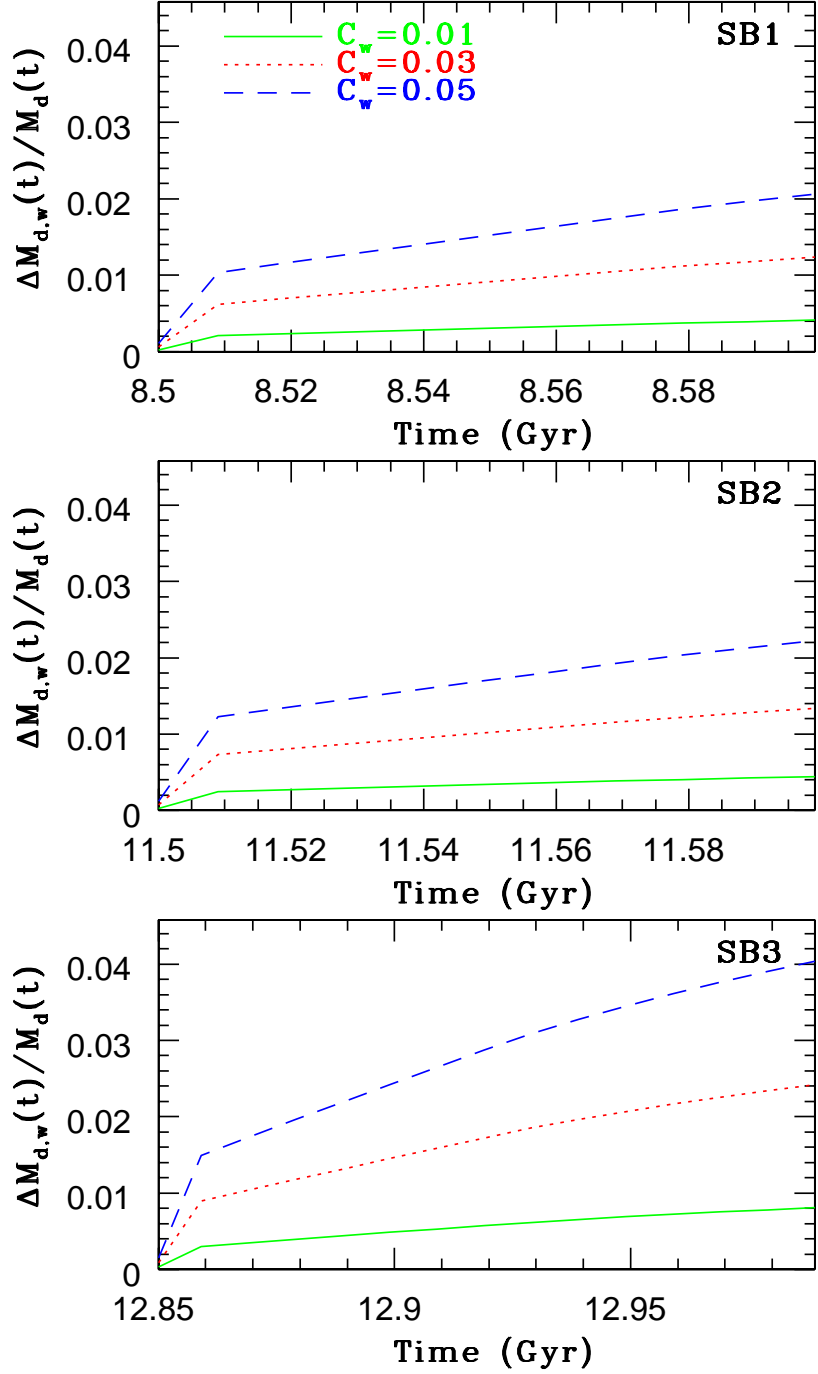}
\figcaption{
Time evolution of $\Delta M_{\rm d, w}(t)/M_{\rm d}(t)$
for the SMC models with $C_{\rm w}=$0.01 (green solid),
$C_{\rm w}=$0.03 (red dotted),
and $C_{\rm w}=$0.05 (blue dashed)
in the three starburst epochs, SB1 (top), SB2 (middle), and SB3 (bottom).
The same value for $C_{\rm w}$ (dust removal coefficient)
is adopted for small and large silicate and carbonaceous  grains 
(i.e., $C_{\rm w, s, Si}=C_{\rm w, l, Si}=C_{\rm w, s, C}=C_{\rm w, l, C}=C_{\rm w}$)
in these SMC models
so that we can clearly demonstrate the physical meaning of $C_{\rm w}$.
The total mass of dust ejected from the SMC due to radiation-driven stellar wind
($\Delta M_{\rm d, w}(t)$) is estimated at each time step  and 
$\Delta M_{\rm d, w}$ normalized  
to the total dust mass ($M_{\rm d}(t)$) at the time step is shown in this figure.
Given the time step width of $10^6$ yr in the present study,
$\Delta M_{\rm d, w}(t)/M_{\rm d}(t)=0.01$ means that the SMC can lose 1\% of its dust
through dust wind in $10^6$ yr.
\label{fig-11}}
\end{figure}

\epsscale{1.0}
\begin{figure}
\plotone{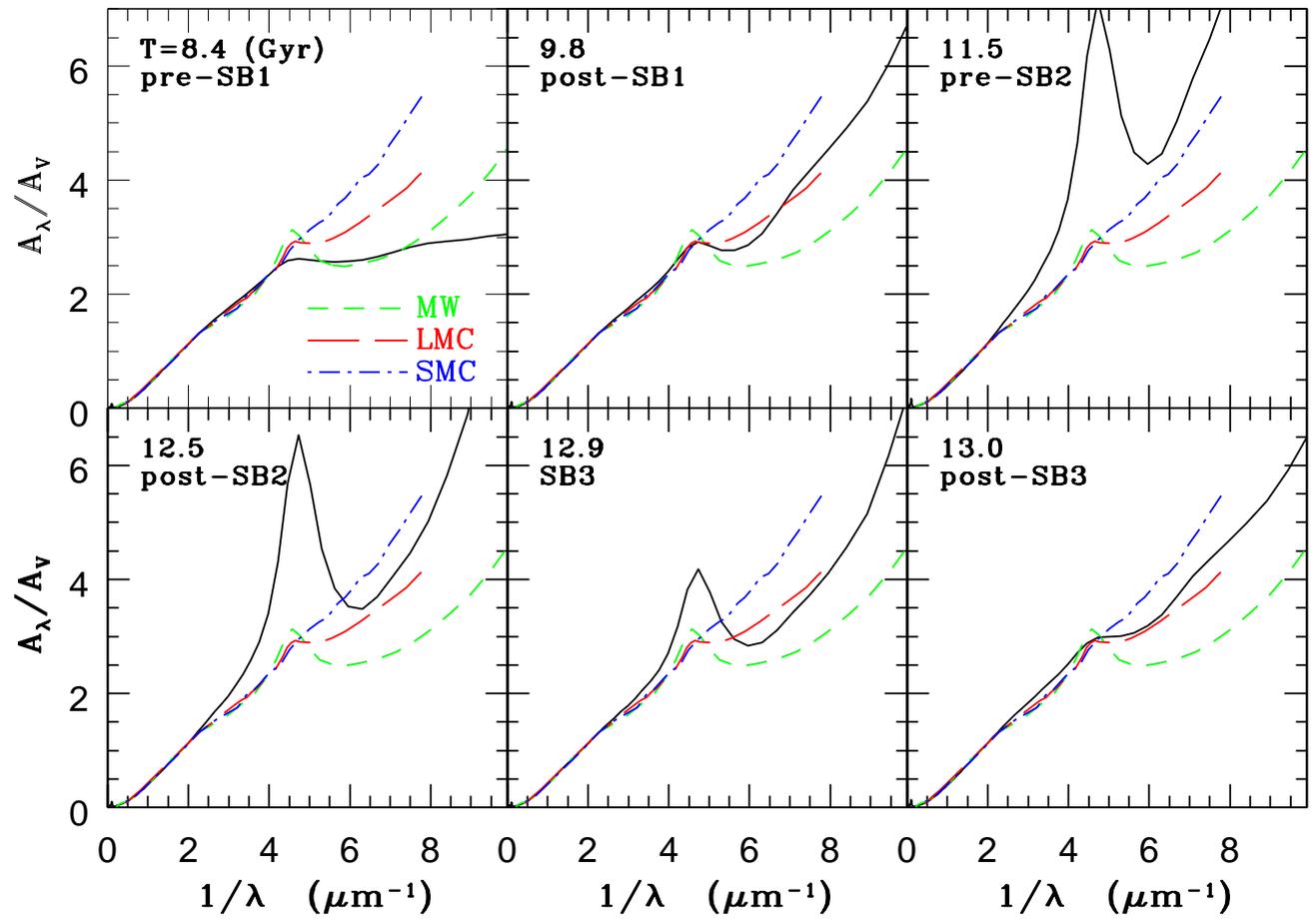}
\figcaption{
Same as Figure 2 but for the SMC model in which the strength of the first starburst 
(SB1) at $T=8.5$ Gyr
is twice as strong as SB1 in the fiducial model S1. In this model, other model parameters
are exactly the same as those of S1. 
\label{fig-12}}
\end{figure}
\clearpage

\epsscale{1.0}
\begin{figure}
\plotone{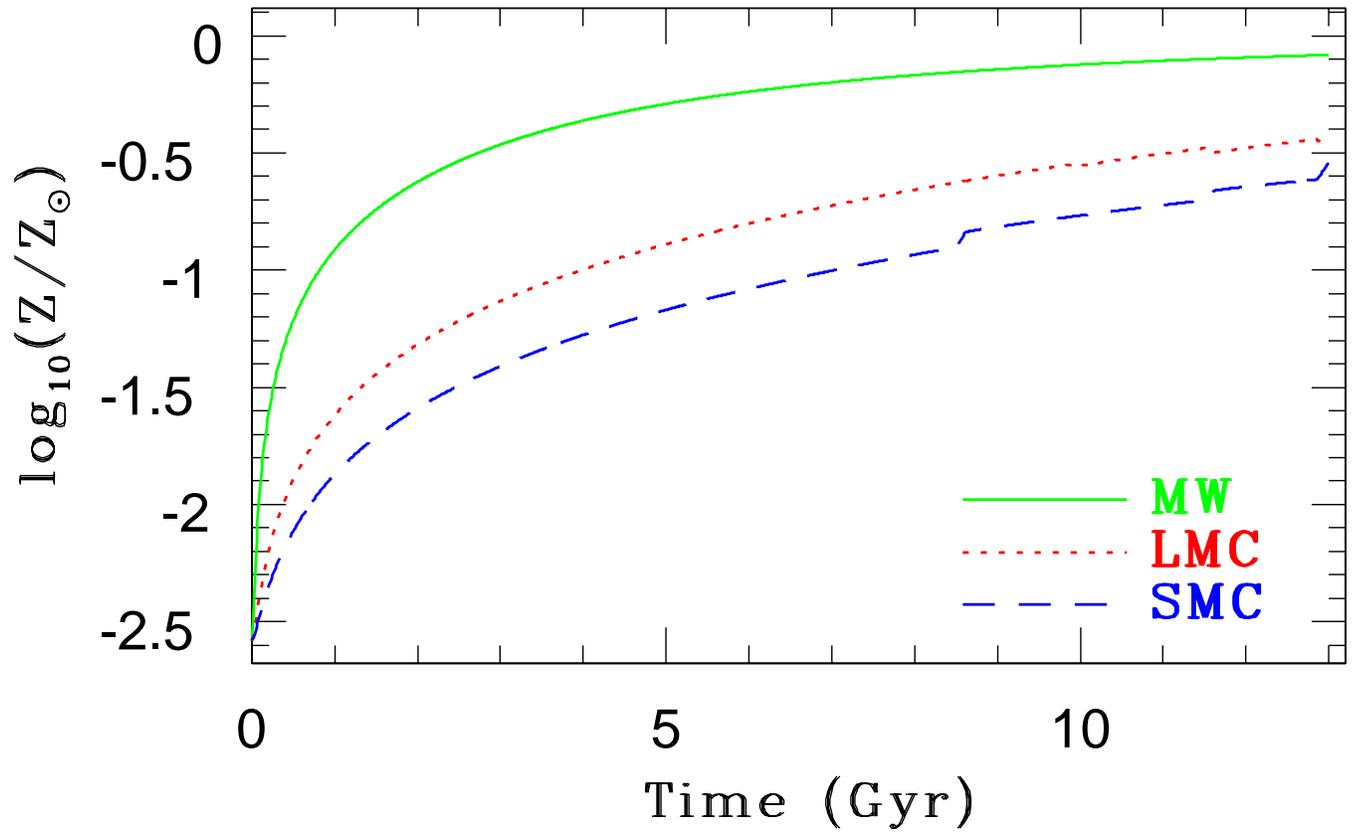}
\figcaption{
The time evolution of metallicities ($Z$ normalized by the solar metallicity $Z_{\odot}$)
for the standard MW (M1; green solid), LMC (L1; red dotted), and SMC (S1; blue dotted)
models. 
\label{fig-13}}
\end{figure}
\clearpage

\clearpage
\appendix
\section{The physical meaning of $C_{\rm w}$}

A key parameter for the time evolution of extinction curves
of galaxies is  $C_{\rm w}$ (dust removal coefficient) in the present study.
Although an adopted value of $C_{\rm w}$ in  a model can indicate whether
dust can be  removed more efficiently in the model than in
other models with different $C_{\rm w}$,
the value of $C_{\rm w}$ (e.g., 0.01)  itself 
does not represent the net rate of dust removal 
(e.g., $0.01 M_{\odot}$ yr$^{-1}$) at each time step in the  model.
It is therefore important for the present study to relate  $C_{\rm w}$ to the
net rate of dust removal.  Figure 11 shows the dust removal rate 
($\Delta M_{\rm d, w}(t)/M_{\rm d}(t)$) at each time step in the three SMC models
with different $C_{\rm w}$. It is clear that (i) the dust removal rate is systematically
higher for the model with larger $C_{\rm w}$ and (ii) it is higher
in later starburst (i.e., higher in SB3 than in SB1).

The dust removal rate of 0.01 at a time step during a starburst in a model
means that 1\% of dust at the time step can be removed from the galaxy 
within the time step width of the model ($10^6$ yr). This furthermore means that
during the starburst with the duration of $10^8$  yr,
the galaxy can lose $\sim$63\% (=$[1-0.99^{100} ] \times 100$\%)  of dust that
it had before the starburst. 
The average $\Delta M_{\rm d, w}(t)/M_{\rm d}(t)$ is 0.0043, 0.013, and 0.022
for the three models with $C_{\rm w}=0.01$, 0.03, and 0.05, respectively. 
Therefore,   $\Delta M_{\rm d, w}(t)/M_{\rm d}(t)$ (per Myr) can correspond roughly to
$0.4C_{\rm w}$ in the present study. 
It should be noted here that even for a constant $C_{\rm w}$, 
$\Delta M_{\rm d, w}(t)/M_{\rm d}(t)$ of a galaxy can significantly change with time
during a starburst
owing to the rapid evolution of the total stellar luminosity of the galaxy 
during the starburst (See Figure 11).
The normalized dust removal rate
($\Delta M_{\rm d, w}(t)/M_{\rm d}(t)$) might be useful for other theoretical models
(e.g., one-zone models and numerical simulations with dust-related physical processes)
that discuss the time evolution of dust in galaxies, particularly when
they include radiation-driven dust wind.

\section{Dependence on the strength of starburst in the SMC}

Although we consider that we have chosen a reasonable and realistic combination
of $C_{\rm sb1}$, $C_{\rm sb2}$, and $C_{\rm sb3}$ for the SMC, 
it could be possible that the present results could depend
on these three parameters.  A particularly interesting question is whether
the SMC can have an extinction curve 
without the 2175 \AA  $\;$ bump yet with the steep rise in the FUV regime
in the early evolution phase if there is a stronger starburst. 
We have accordingly investigated the SMC models in which only  $C_{\rm sb1}$ is different
from the fiducial SMC model S1 so that we can address this issue.
Figure 12 describes  the time evolution of the extinction curve
in the model with $C_{\rm sb1}=0.1$ (instead of 0.05 in the fiducial SMC model),
which means that the first starburst is as strong as the third one in the model.

Clearly, the extinction curves in the early phase ($T\le 9.8$ Gyr) of this model
are not so similar to the observed one of the present SMC, 
and this is true for the model even with $C_{\rm sb1}=0.2$. These results accordingly
suggest that an extinction curve similar to that observed 
in the present SMC can not be achieved
in the present models at  early evolutionary  phases  by changing the strength of SB1 alone.
These also imply that the SMC might have established its characteristic extinction
curve quite recently. 
The final extinction curve in this model looks similar to the observed one for the SMC,
but the degree of such similarity is higher in S1 than in this model,
which means that the present fiducial model S1 chooses a more realistic  parameter value
than this model 
with $C_{\rm sb1}=C_{\rm sb3}=0.1$ does. It is interesting that this model
shows a  more conspicuous  2175 \AA  $\;$ bump 
at $T=11.5$ Gyr and 12.5 Gyr. It is confirmed that the observed  SMC extinction curve
can not be reproduced well 
in the SB2 phase of the SMC models with higher
$C_{\rm sb2}$ (=0.1). 

\section{Metallicity evolution}

The time evolution of metallicities ($Z$)  of the MCs and the Galaxy
has been already investigated
in detail by Tsujimoto \& Bekki (2009), Tsujimoto et al. (2010), and  BT12. 
However, the models adopted in these studies are different from the present one
in the sense that they did not include dust-related physical processes
(e.g., dust growth). Therefore, it is instructive for the present study
to present the results of metallicity  evolution in the new models.
Figure 13 shows that the global trends of the metallicity evolution in
the present three galaxy models (S1, L1, and M1) are very similar to those derived in our
previous studies. However, the final metallicities at $T=13$ Gyr are slightly smaller 
than the observed values of the galaxies: $Z=0.0046$, 0.0058, and 0.013 for
S1, L1, and M1 models, respectively. These  $Z$ slightly smaller than
the observed values are  due largely to
the more efficient 
loss of dust and metals through SN and dust winds in the present new models.
The models with less efficient SN and dust winds can show higher $Z$ that are
closer to the observed values.

\clearpage

\end{document}